\documentclass[a4paper,10pt]{article}
\usepackage[T1]{fontenc}      
\usepackage[utf8]{inputenc}   
\usepackage[english]{babel}   
\usepackage{url}              
\usepackage{bm}
\usepackage{amsmath}
\usepackage{amsfonts}
\usepackage{graphicx}
\usepackage[]{appendix}
\usepackage{listings,lstautogobble}

\usepackage{tabularx}
\usepackage{pdflscape}
\usepackage{longtable}
\usepackage{bigints}
\usepackage{amsmath}
\usepackage{amssymb}
\usepackage{float}
\usepackage[footnotesize,bf]{caption2}
\usepackage{bm}
\usepackage{geometry}
\usepackage{etoolbox}
\usepackage{amsthm}
\usepackage{amsmath}
\usepackage{xcolor}
\usepackage{algorithm}
\usepackage{algpseudocode}
\usepackage{bm}

\usepackage{booktabs}
\usepackage{threeparttable}
\usepackage{float}

\usepackage{geometry}
\usepackage{graphicx}
\graphicspath{{img/}}

\usepackage{scalerel}
\usepackage{xcolor}

\usepackage{multirow}

\usepackage{algpseudocode}

\definecolor{blucite}{RGB}{12,127,172}

\usepackage[authoryear]{natbib}
\usepackage[colorlinks, citecolor=blucite]{hyperref}
\usepackage{doi}

\title{\LARGE \textbf{Impact of Near-Positivity Violations on IPTW-Estimated Marginal Structural Survival Models With Time-Dependent Confounding}\\\vspace{0.4cm}}
\author{\Large  Marta Spreafico\\
	\normalsize \href{mailto:m.spreafico@math.leidenuniv.nl}{\normalsize \texttt{m.spreafico@math.leidenuniv.nl}}\\
	\vspace{-2mm}\\
		\footnotesize \textit{Mathematical Institute, Leiden University, Leiden 2333 CC, The Netherlands}\\
	\footnotesize \textit{Department of Biomedical Data Sciences, Leiden University Medical Center, Leiden 2333 ZA, The Netherlands}}
\date{January 13, 2025}

\begin{document}
\normalsize
\maketitle

\vspace{1mm}
\noindent \textbf{Disclaimer}:  This preprint is the submitted version of an article accepted for publication in \textit{Biometrical Journal}. The final, peer-reviewed version is available at doi: \href{https://onlinelibrary.wiley.com/doi/10.1002/bimj.70093}{10.1002/bimj.70093}.
\vspace{2mm}
		
\begin{abstract}
	\small
	In longitudinal observational studies, marginal structural models (MSMs) are a class of causal models used to analyse the effect of an exposure on the (time-to-event) outcome of interest, while accounting for exposure-affected time-dependent confounding. In the applied literature, inverse probability of treatment weighting (IPTW) has been widely adopted to estimate MSMs. 
	An essential assumption for IPTW-based MSMs is the positivity assumption, which ensures that, for any combination of measured confounders among individuals, there is a non-zero probability of receiving each possible treatment strategy. 
	Positivity is crucial for valid causal inference through IPTW-based MSMs, but is often overlooked compared to confounding bias.
	Positivity violations may also arise due to randomness, in situations where the assignment to a specific treatment is theoretically possible but is either absent or rarely observed in the data, leading to near violations. 
	These situations are common in practical applications, particularly when the sample size is small, and they pose significant challenges for causal inference.
	This study investigates the impact of near-positivity violations on estimates from IPTW-based MSMs in survival analysis. 
	Two algorithms are proposed for simulating longitudinal data from hazard-MSMs, accommodating near-positivity violations, a time-varying binary exposure, and a time-to-event outcome.
	Cases of near-positivity violations, where remaining unexposed is rare within certain confounder levels, are analysed across various scenarios and weight truncation (WT) strategies. 
	This work aims to serve as a critical warning against overlooking the positivity assumption or naively applying WT in causal studies using longitudinal observational data and IPTW. \\
	\vspace{2mm}
\end{abstract}

\vspace{1mm}
\normalsize

\maketitle                   

\section{Introduction}\label{s:intro}

In longitudinal observational studies, exposure-affected time-varying confounding represents a specific challenge for estimating the effect of a treatment on the (time-to-event) outcome of interest, as standard analyses fail to give consistent estimators \citep{daniel2013,clare2019}. In the past decades considerable progresses have been made in developing causal methods for analysing such complex longitudinal data.
Marginal Structural Models (MSMs) were introduced as a powerful method for confounding control in longitudinal studies \citep{daniel2013,robins2000,Hernan2000,ravani2017}, alternatively to structural nested models \citep{robins1992}. MSMs are models for the \textit{potential} or \textit{counterfactual} outcome that individuals would have experienced if they had received a particular treatment or exposure value.
This study focuses on counterfactual time-to-event outcomes by considering marginal structural hazard models (hazard-MSM) or a discrete-time analogue. 
The parameters of a MSM can be consistently estimated through various methods, including Inverse Probability of Treatment Weighting (IPTW) estimators, G-computation, or doubly robust methods \citep{daniel2013,clare2019,robins2000,Van_der_Laan2016-gj,cibook2020,Gabriel2023}.
Despite less robust, IPTW-based MSMs have largely been adopted in the applied literature, especially in epidemiology and medicine, due to its simplicity in both implementation and interpretation \citep{clare2019}.
IPTW-based MSMs require the correct specification of the exposure model conditional on confounders (i.e., the \textit{weighting model}) and special attention to the identifiability assumptions of consistency, no unmeasured confounding, and positivity \citep{cibook2020,colehernan2008,cole2009epi,ravani2017}. This work focuses on the latter, which is often overlooked compared to confounding bias.

Positivity holds if, for any combination of the measured confounders occurring among individuals in the population, there is a non-zero (i.e., positive) probability of receiving every level of the exposure possible under the target treatment strategies to be compared \citep{colehernan2008,cibook2020}.
While less well-recognized than bias due to incomplete control of confounding, violations of the positivity assumption can increase both the variance and bias of causal effect estimates \citep{petersen2012,Leger2022}.
Positivity violations can occur in two situations  \citep{Zhu2021}.
\textit{Strict} (or \textit{theoretical}) violations occur when certain treatment levels are impossible for specific subgroups of subjects. For example, if a certain treatment $a$ is never given to individuals with severe comorbidities, then the causal effect of $a$ cannot be estimated for that subgroup; the analysis should therefore focus only on individuals without severe comorbidities.
Even in the absence of structural zeros, \textit{random} zeros may occur by chance due to small sample sizes  or highly stratified data by numerous confounders.
\textit{Near} (or \textit{practical}) violations refer to situations where the assignment to a specific treatment is always theoretically possible but is not (or rarely) observed in the data due to randomness. Sampling variability may indeed result in subjects having a near-zero probability of being exposed (or unexposed) for certain combinations of covariate values.  
These situations are common in practical applications, particularly when the sample size is small, and they pose significant challenges for causal inference. While treatment remains technically possible within a subgroup, its rarity makes reliable estimation difficult, particularly when using IPTW. Extremely small probabilities lead to extreme weights, which can destabilize the analysis and inflate variance. To address these challenges and stabilize the variance of estimates, Weight Truncation (WT) is often applied in IPTW to down-weight observations in regions where near-violations occur  \citep{colehernan2008,Xiao2013,Zhu2021}. However, if applied inappropriately, this technique could result in excessive truncation, which may introduce bias into the estimates.

In the literature, research studies on positivity violations have been previously carried out in a pedagogical manner by using real data to illustrate how incorrect inference occurs in estimating MSMs when positivity is violated \citep{Mortimer2005,Bembom2007,colehernan2008,Zhu2021,Rudolph2022,Zhu2023}. Findings from different studies agreed that positivity violations have a more severe impact for the IPTW-estimator than other causal estimators. However, when using real data, disentangling the effect of positivity violations from other sources of bias is typically not possible: important confounders may be undetected or unmeasured and the fulfilment of the remaining assumptions underlying the IPTW-estimator is generally difficult to ascertain. Moreover, real data do not allow us to design scenarios that could be of interest, such as studying performance under different sample sizes. To overcome these limitations, other investigations were conducted more systematically by setting up simulation studies \citep{Neugebauer2005,Wang2006,Naimi2011,petersen2012,Leger2022}. Results confirmed that under positivity violations IPTW-estimator performs worse than other methods, becoming very unstable and exhibiting high variability.
However, these studies were limited to assessing the causal effect of a treatment assigned either at a single time point or twice.

This study investigates the impact of near-positivity violations on the performance of IPTW-estimated MSMs in longitudinal survival contexts with time-varying confounding using a simulation-based approach. 
No systematic simulation studies currently exist in this framework, largely due to the challenges of simulating longitudinal survival data under conditions of both exposure-affected time-varying confounding and near-positivity violations. 
To address this gap, two algorithms are proposed for simulating longitudinal data from hazard-MSMs, accommodating near-positivity violations, a time-varying binary exposure, and a survival outcome. These methods build on the works of \cite{Havercroft2012} and \cite{Keogh2021}.
Two simulation studies analysing cases of near-positivity violations, where remaining unexposed is rare within certain levels of confounders, through various scenarios and WT strategies are performed.

This study aims to highlight the critical importance of carefully considering the positivity assumption in causal studies that utilize longitudinal observational data and IPTW. The final purpose is to warn against the risks of underestimating the assumption's significance or uncritically applying WT methods.
This is fundamental given that the past several decades have seen an exponential growth in causal
inference approaches and their applications to observational data \citep{Hammerton2021,Mitra2022,olier2023}, including emerging areas such as target trial emulations  \citep{hernan2016,hernan2021}, prediction modelling under hypothetical interventions \citep{vanGeloven2020,Lin2021,Keogh2024}, or causal machine learning \citep{Feuerriegel2024,Moccia2024}.

This study is organized as follow. Section \ref{sec:methods} briefly recalls the notation, MSMs for survival outcomes, and IPTW. Section \ref{sec:simu} explains the proposed mechanism to enforce positivity violations in algorithms to simulate longitudinal data from MSMs. Sections \ref{sec:simuI} and \ref{sec:simuII} present the two simulation studies. Section \ref{sec:discussion} finally discusses the findings and provide recommendations.
Statistical analyses were performed in the \texttt{R} software environment \citep{Rsoftware}. Source code is available at {\url{https://github.com/mspreafico/PosViolMSM}}. 

\section{Marginal structural models for potential survival outcomes}\label{sec:methods}

\subsection{Notation}\label{sec:methods:not}
Let us consider a set of $i=1,\dots,n$ subjects and a set of regular visits $k=0,1,\dots,K$ performed at times $q_0<q_1<\dots<q_K$ (assumed to be the same for everybody). Each subject undergoes each visit up until event time $T^*_i = \min(C_i,T_i)$, i.e., the earlier of the time of the actual event of interest $T_i$ and the administrative censoring time $C_i$.
At each visit $k$, if $T_i \ge q_k$, a binary treatment status $A_{i,k} \in \{0,1\}$ (unexposed vs exposed; control vs treatment) and a set of time-dependent covariates $L_{i,k}$ are observed. A bar over a time-dependent variable indicates the history, that is $\bar{\boldsymbol{A}}_{i,k} = \left(A_{i,0},\dots,A_{i,k}\right)$ and $\bar{\boldsymbol{L}}_{i,k} = \left(L_{i,0},\dots,L_{i,k}\right)$. Finally, the binary failure indicator process is denoted by $Y_{i,k+1}$, where $Y_{i,k+1}=1$ if subject $i$ has failed (e.g., died) in period $(q_k; q_{k+1}]$, i.e., $q_k < T_i \le q_{k+1}$, or $Y_{i,k+1}=0$ otherwise. 

\subsection{Marginal structural models for counterfactual hazard rates}\label{sec:methods:MSMs}
Marginal structural hazards models (hazard-MSMs) are a class of causal models that focus on \textit{counterfactual} time-to-event variables \citep{Hernan2000,robins2000,cibook2020}. These variables represent the time at which an event would have been observed had a patient been administered a specific exposure strategy $\bar{\boldsymbol{a}} = \left(a_0,a_1,\dots,a_K\right)$ with $a_k \in \{0,1\}$ for all $k$. Vector $\bar{\boldsymbol{a}}$ might differ from the actual treatment received $\bar{\boldsymbol{A}}_i = \bar{\boldsymbol{A}}_{i,K} = \left(A_{i,0},\dots,A_{i,K}\right)$.
The \textit{counterfactual event time} that would be observed in a subject under complete exposure history $\bar{\boldsymbol{a}}$ is denoted by $T^{\bar{\boldsymbol{a}}}$. Hazard-MSMs hence model the counterfactual hazard rate:
\begin{equation*}
	\lambda^{\bar{\boldsymbol{a}}}(t) =\lim_{\Delta t \to 0} \frac{P(t\leq T^{\bar{\boldsymbol{a}}} < t+\Delta t \mid T^{\bar{\boldsymbol{a}}}\geq t)}{\Delta t}. 
\end{equation*}

In case of discrete-time hazard of failure, a \textit{marginal structural logistic regression model} (\textit{logit-MSM}) can be assumed to model 
the counterfactual probability of failure in
a single interval $(q_k,q_{k+1}]$, given survival up to $q_k$. The \textit{logit-MSM} for the counterfactual hazard at visit $k$ is defined as follows:
\begin{equation}\label{eq:MSMlogit}
	\lambda^{\bar{\boldsymbol{a}}}_k =  \Pr\left(Y^{\bar{\boldsymbol{a}}}_{k+1}=1 \mid Y^{\bar{\boldsymbol{a}}}_{k}=0\right) = \text{logit}^{-1}\left[ \tilde{\gamma}_{0} + g\left(\tilde{\boldsymbol{\gamma}}_A;
	\bar{\boldsymbol{a}}_{k}\right)\right],
\end{equation}
where $\bar{\boldsymbol{a}}$ is the the complete treatment strategy, $Y^{\bar{\boldsymbol{a}}}_{k}$ is the counterfactual event indicator at visit $k$, $g(\cdot)$ is a function (to be specified) of the treatment strategy history up to visit $k$ (denoted by $\bar{\boldsymbol{a}}_{k}$), and $(\tilde{\gamma}_{0},\tilde{\boldsymbol{\gamma}}_A)$ is the vector of log odd ratios, with $\tilde{\gamma}_{0}$ as the intercept.

In the context of continuos-time hazard, the \textit{marginal structural Cox proportional hazard model} (\textit{Cox-MSM}) form is often assumed to model the counterfactual hazard at time $t$ given treatment history $\bar{\boldsymbol{a}}$:
\begin{equation} \label{eq:MSMcox}
	\lambda^{\bar{\boldsymbol{a}}}(t) = \lambda_{0}(t) \exp\left\{g\left(\tilde{\boldsymbol{\beta}}_A;
	\bar{\boldsymbol{a}}_{\lfloor{t}\rfloor}\right)\right\},
\end{equation}
where $\lambda_{0}(t)$ is the baseline hazard function, $\bar{\boldsymbol{a}}_{\lfloor{t}\rfloor}$ denotes treatment pattern up to the most recent visit prior to time $t$ (i.e., $\lfloor{t}\rfloor = \max_{k\le t} k$), $g(\cdot)$ is a function (to be specified) of treatment pattern $\bar{\boldsymbol{a}}_{\lfloor{t}\rfloor}$, and $\tilde{\boldsymbol{\beta}}_A$ is a vector of log hazard ratios.
An alternative often used is the \textit{marginal structural Aalen's additive hazard model} (\textit{Aalen-MSM}), defined as
\begin{equation}\label{eq:MSMaalen}
	\lambda^{\bar{\boldsymbol{a}}}(t) = \tilde{\alpha}_{0}(t)+ g\left(\tilde{\boldsymbol{\alpha}}_A(t);\bar{\boldsymbol{a}}_{\lfloor{t}\rfloor}\right)
\end{equation}
where $\tilde{\alpha}_{0}(t)$ is the baseline hazard at time $t$, $\bar{\boldsymbol{a}}_{\lfloor{t}\rfloor}$ denotes treatment pattern up to the most recent visit prior to time $t$, $g(\cdot)$ is a function (to be specified) of treatment pattern $\bar{\boldsymbol{a}}_{\lfloor{t}\rfloor}$, and $\tilde{\boldsymbol{\alpha}}_A(t)$ is the vector of coefficients at time $t$. 

In the hazard-MSMs \eqref{eq:MSMlogit}-\eqref{eq:MSMaalen}, the function $g(\cdot)$ combines information from the treatment pattern up to the most recent visit $k$ prior to time $t$. Depending on the desired information provided in $g(\cdot)$, the hazard at time $t$ (or visit $k$) can thus assume different forms.
Table \ref{tab:gform} shows examples of three forms of the treatment pattern function $g(\cdot)$, specifically: (i) the current level of treatment, (ii) the duration of treatment, or (iii) the history of treatment up to time $t$ through the main effect terms for treatment at each visit. Any other desired form can be alternatively specified.

\begin{table}[htb]
	\begin{center}
		\caption{Examples of treatment-pattern forms for function $g(\cdot)$ in Equations \eqref{eq:MSMlogit} to \eqref{eq:MSMaalen}.}	\label{tab:gform}
		\begin{tabular}{lccc}
			\hline\noalign{\smallskip}
			\textbf{Form of treatment pattern} & \textbf{Logit-MSM}  & \textbf{Cox-MSM}  & {\textbf{Aalen-MSM}}  \\
			Function $g(\cdot)$ & $g\left(\tilde{\boldsymbol{\gamma}}_A;
			\bar{\boldsymbol{a}}_{k}\right)$ in \eqref{eq:MSMlogit} & $g\left(\tilde{\boldsymbol{\beta}}_A;\bar{\boldsymbol{a}}_{\lfloor{t}\rfloor}\right)$ in \eqref{eq:MSMcox} & $g\left(\tilde{\boldsymbol{\alpha}}_A(t);\bar{\boldsymbol{a}}_{\lfloor{t}\rfloor}\right)$ in \eqref{eq:MSMaalen}\\
			\noalign{\smallskip}\hline\noalign{\smallskip}
			Current level of treatment & $\tilde{\gamma}_A \cdot a_{k}$ & $ \tilde{\beta}_{A} \cdot a_{\lfloor{t}\rfloor}$ & $\tilde{\alpha}_{A}(t) \cdot a_{\lfloor{t}\rfloor}$\\
			\noalign{\smallskip}\hline\noalign{\smallskip}
			Duration of treatment & $ \tilde{\gamma}_A \cdot \sum_{j=0}^{{k}} a_{{k}-j}$ & $\tilde{\beta}_A \cdot \sum_{j=0}^{{\lfloor{t}\rfloor}} a_{{\lfloor{t}\rfloor}-j}$ & $\tilde{\alpha}_A(t) \cdot \sum_{j=0}^{{\lfloor{t}\rfloor}} a_{{\lfloor{t}\rfloor}-j}$ \\
			\noalign{\smallskip}\hline\noalign{\smallskip}
			Main effect terms at each visit & $\sum_{j=0}^{{k}} \tilde{\gamma}_{Aj}  \cdot a_{{k}-j}$ & $\sum_{j=0}^{{\lfloor{t}\rfloor}} \tilde{\beta}_{Aj} \cdot a_{{\lfloor{t}\rfloor}-j}$ & $\sum_{j=0}^{{\lfloor{t}\rfloor}} \tilde{\alpha}_{Aj}(t) \cdot a_{{\lfloor{t}\rfloor}-j}$\\
			\hline
		\end{tabular}	
	\end{center}
\end{table}

\subsubsection{Hazard-based estimands and marginal survival probabilities}\label{sec:methods:MSMs:surv}
The logit-MSM estimates log odd ratios $\tilde{\boldsymbol{\gamma}}_A$, the Cox-MSM the log hazard ratios $\tilde{\boldsymbol{\beta}}_A$, and the Aalen-MSM the cumulative regression coefficients $\int_{0}^t \tilde{\boldsymbol{\alpha}}_A(s) ds$. Since hazard-based estimands may not have a straightforward interpretation, estimates from the MSMs are typically transformed into estimates for an interpretable causal estimand \citep{Hernan2010haz,Martinussen2020,Keogh2021,Didelez2022}. One example is comparing the marginal survival probabilities at time $t$, i.e., $S^{\bar{\boldsymbol{a}}}(t) = \Pr(T^{\bar{\boldsymbol{a}}} >  t)$, for \textit{always treated} $\bar{\boldsymbol{a}}= \boldsymbol{1} = (1,\dots,1)$ (i.e., sustained use of the treatment) versus \textit{never treated} $\bar{\boldsymbol{a}}= \boldsymbol{0} = (0,\dots,0)$ (i.e., sustained non-use of the treatment), or evaluating the marginal risk difference between them.
The marginal survival probability at time $t$ under treatment history $\bar{\boldsymbol{a}}$ can be computed based on the different hazard forms:
\begin{itemize}
	\item [(i)] for the logit-MSMs in \eqref{eq:MSMlogit} is given by
	\begin{equation}\label{eq:MSMlogit:surv}
		S^{\bar{\boldsymbol{a}}}(t) 
		= \prod_{k \leq t}\left(1-\lambda^{\bar{\boldsymbol{a}}}_k\right) = \prod_{k \leq t}\left(1- \text{logit}^{-1}\left[ \tilde{\gamma}_{0} + g\left(\tilde{\boldsymbol{\gamma}}_A;
		\bar{\boldsymbol{a}}_{k}\right)\right] \right);
	\end{equation}
	\item [(i)] for the Cox-MSM in \eqref{eq:MSMcox} is given by
	\begin{equation}\label{eq:MSMcox:surv}
		S^{\bar{\boldsymbol{a}}}(t) 
		= 
		\exp\Bigg(-e^{g(\tilde{\boldsymbol{\beta}}_A;
			a_{0})} \int_{0}^{1}\lambda_{0}(s)\,ds -e^{g(\tilde{\boldsymbol{\beta}}_A;
			\bar{\boldsymbol{a}}_{1})} \int_{1}^{2}\lambda_{0}(s)\,ds -\cdots - e^{g(\tilde{\boldsymbol{\beta}}_A;
			\bar{\boldsymbol{a}}_{\lfloor{t}\rfloor})} \int_{\lfloor{t}\rfloor}^{t}\lambda_{0}(s)\,ds 
		\Bigg);
	\end{equation}
	\item [(iii)] for the Aalen-MSM in \eqref{eq:MSMaalen} is given by
	\begin{equation}\label{eq:MSMaalen:surv}
		S^{\bar{\boldsymbol{a}}}(t) 
		= 
		\exp\left(-\int_{0}^{t}\tilde{\alpha}_{0}(s)\,ds 
		-\int_{0}^{1}g(\tilde{\boldsymbol{\alpha}}_{A}(s);
		a_{0})\,ds -\int_{1}^{2}g(\tilde{\boldsymbol{\alpha}}_{A}(s);
		\bar{\boldsymbol{a}}_{1})\,ds -
		\cdots
		-\int_{\lfloor{t}\rfloor}^{t}g(\tilde{\boldsymbol{\alpha}}_{A}(s);
		\bar{\boldsymbol{a}}_{\lfloor{t}\rfloor})\,ds 
		\right).
	\end{equation}
\end{itemize}

\subsection{Inverse Probability of Treatment Weighting (IPTW)}\label{sec:methods:IPTW}
In the presence of confounders, MSMs can be estimated from the observed data by applying a technique called inverse probability of treatment weighting (IPTW) \citep{cibook2020}. IPTW involves weighting the contribution of each subject $i$ by the inverse of the probability of receiving their actual exposure level given their confounding covariates. This process creates a pseudo-population where confounding is no longer present.
To optimize the variance estimation, stabilized (or standardized) weights are usually preferred \citep{robins2000,Hernan2000,cibook2020,Leger2022}.
The stabilized weight for subject $i$ at time $t$ is defined as
\begin{equation}\label{eq:sw}
	sw_{i}(t) = \prod_{k=0}^{\lfloor t \rfloor} \frac{\Pr\left(A_{i,k} \mid \boldsymbol{\bar{A}}_{i, k-1},T_i \ge q_k\right)}{\Pr\left(A_{i, k}\mid\boldsymbol{\bar{A}}_{i,k-1},\boldsymbol{\bar{L}}_{i,k},T_i \ge q_k\right)},
\end{equation}
where $\lfloor{t}\rfloor = \max_{k\le t} k$ is the largest visit-time prior to $t$, and ${A}_{-1}$ is defined to be 0. In the pseudo-population thus created, the effects of time-dependent confounding are balanced, so association in hazard regression models is causation \citep{cibook2020}.

Even when standardized, weights $sw_i(t)$ can largely inflate for a subject $i$ concerned by near-positivity violation: when the denominator probabilities are very close to zero, weights become extremely large.
In such cases, the common approach is to consider truncated stabilized weights $\tilde{sw}_i(t)$ obtained by truncating the lowest and the highest estimations by the 1st and 99th (1-99) percentiles, or alternatively by narrower truncations, such as the 2.5-97.5, 5-95, or 10-90 percentiles \citep{colehernan2008,Xiao2013,Zhu2021}.

\subsection{Simulating longitudinal survival data from marginal structural hazard models}\label{sec:methods:sim}

Even when positivity holds, simulating longitudinal data when the model of interest is a model for potential outcomes, as for MSMs, is generally not straightforward \citep{Evans2023}. The main challenges consist in: (i) replicating the complex dynamics of time-varying confounding; (ii) generating data in such a way that the model of interest is correctly specified; and (iii) in case of survival or other non-collapsible models \citep{Robinson1991,Didelez2022}, reconciling the MSM with the conditional model used in Monte Carlo studies to generate the data. For these reasons, only a few methods for simulating data from hazard-MSMs have been published in the literature. These methods impose restrictions on the data-generating mechanisms to address the issues mentioned, allowing for accurate simulation of longitudinal data from pre-specified hazard-MSMs. 

\cite{Xiao2010} and \cite{Young2010} first introduced two approaches for simulating from Cox-MSMs. \cite{Havercroft2012} and \cite{Young2014} outlined how to simulate in a discrete-time setting, from marginal structural logistic regression models and discrete-time Cox-MSMs, respectively. By finely discretizing time, these algorithms can be adapted to simulate from a continuous-time Cox-MSM.  \cite{Keogh2021} introduced a method to simulate from a marginal structural Aalen's additive hazard model that resulted in a less-restrictive generating mechanism. Recently, \cite{Seaman2024} outlined how to simulate from various hazard-MSMs that condition the hazard on baseline covariates.

\begin{figure}[h!]
	\begin{center}
		\includegraphics[width=0.9\textwidth]{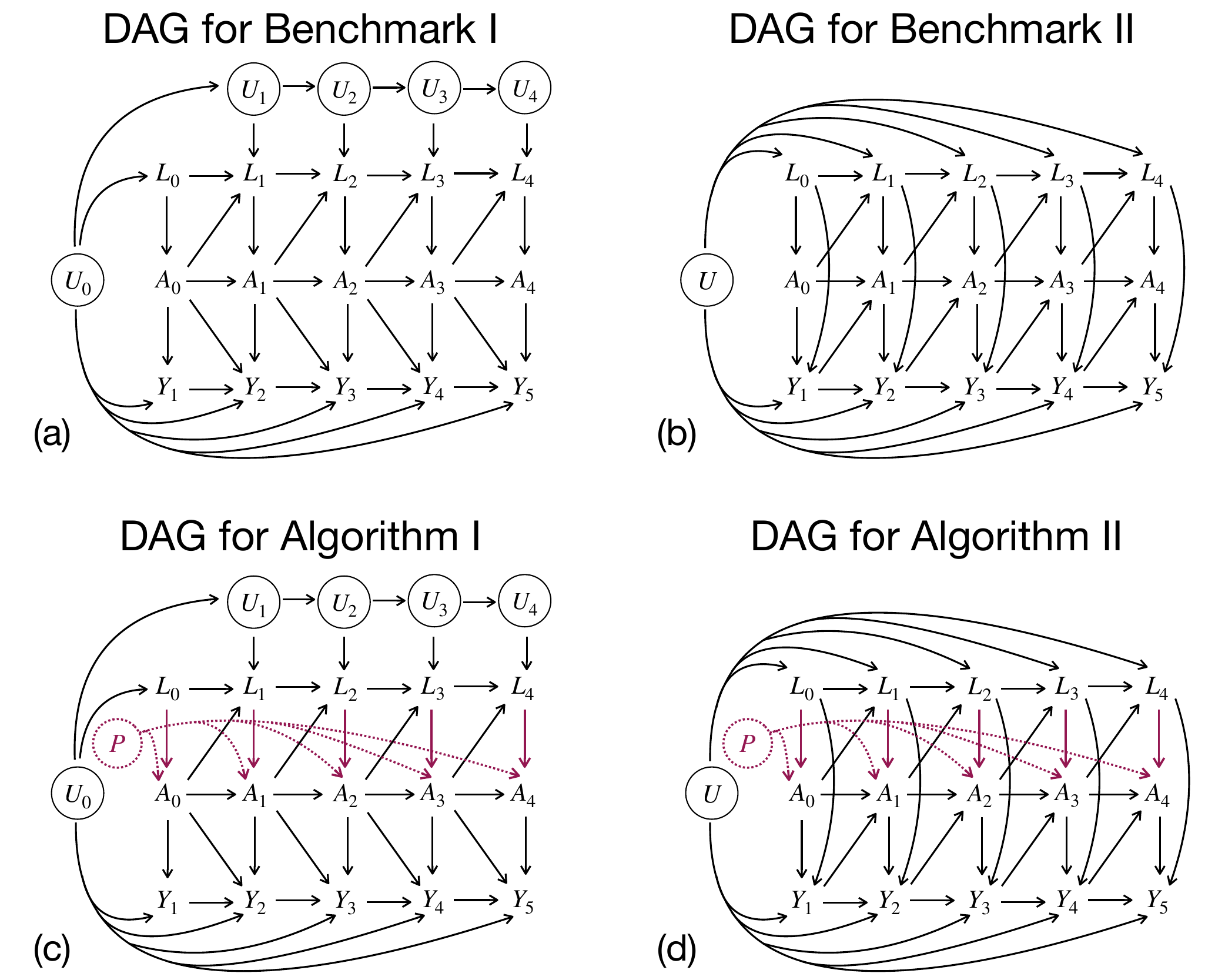}
		\caption{Top: Directed acyclic graphs (DAGs) illustrating the temporal causal relationships between variables in the data-generating mechanisms proposed by (a) \cite{Havercroft2012} and (b) \cite{Keogh2021}, i.e., Benchmarks I and II. Bottom: DAGs illustrating the temporal causal relationships between variables in the proposed data-generating mechanisms, i.e., (c) Algorithm I and (d) Algorithm II.}
		\label{fig:dags}
	\end{center}
\end{figure}

Among these, the approaches by \cite{Havercroft2012} and \cite{Keogh2021} present similar structures of the temporal causal relationships between variables.
The directed acyclic graphs (DAGs) in top panels of Figure \ref{fig:dags} display for both cases the assumed data structure and inform which variables, measured at which time points, are confounders of the association between treatment at a given time point and the outcome. Both DAGs are illustrated in discrete-time for a follow-up with $k=0,\dots,K$ visits. Variables are assumed to be constant between visits.  By imagining splitting the time intervals between successive visits into smaller and smaller intervals, both structures approach the continuous-time setting. 
For the current study, these two approaches are used as ``truth'' benchmarks for cases where longitudinal data generation from the desired hazard-MSM has already been demonstrated and positivity assumption is valid.

\section{Simulating longitudinal survival data with random positivity violations}\label{sec:simu}
To impose near-positivity violations within a data-generating mechanism, certain treatment levels (e.g., exposure or non-exposure in binary treatments) may become unobservable (though theoretically possible) for specific subgroups defined by confounders, due to randomness.
Suppose the interest is on the subgroup of subjects presenting a poor health condition.
Let us assume that the subgroup of subjects presenting a poor health condition at visit $k$ is determined by a range of values $\mathcal{I}_{\tau}$ of the confounder $L_{i,k}$.
Near-positivity violations occur when the probability of remaining unexposed (or being exposed), given a poor health condition, is very close to zero (or approaches one):
\begin{align}\label{eq:nearViol}
	\Pr\left(A_{i,k} = 0 \mid L_{i,k} = l_{i,k}\in \mathcal{I}_{\tau}, \bar{\boldsymbol{L}}_{i,k-1} =\bar{\boldsymbol{l}}_{k-1}, \bar{\boldsymbol{A}}_{i,k-1} =\bar{\boldsymbol{a}}_{k-1},T_i \ge q_k\right) & \approx 0,\\
	\Pr\left(A_{i,k} = 1 \mid L_{i,k} = l_{i,k}\in \mathcal{I}_{\tau}, \bar{\boldsymbol{L}}_{i,k-1} =\bar{\boldsymbol{l}}_{k-1}, \bar{\boldsymbol{A}}_{i,k-1} =\bar{\boldsymbol{a}}_{k-1},T_i \ge q_k\right) & \approx 1, \nonumber
\end{align}
for all $\bar{\boldsymbol{a}}_{k-1}, \bar{\boldsymbol{l}}_{k-1}$ when $\Pr\left(\bar{\boldsymbol{A}}_{i,k-1} =\bar{\boldsymbol{a}}_{k-1},\bar{\boldsymbol{L}}_{i,k} =\bar{\boldsymbol{l}}_{k},T_i \ge q_k\right) \neq 0$. 
This happens when remaining unexposed to treatment at visit $k$ is rarely observable for subjects in the poor health subgroup.

Once defined the poor health subgroup of interest $\mathcal{I}_{\tau}$, violations occurring by chance can be introduced in a data generating mechanism by considering (i) a latent individual propensity $P_i$ to exposure given a poor health condition, and (iii) an exposure cut-off $\pi \in [0,1]$. 
For each subject $i$, a random uniform variable $P_i$ is generated on the interval $[0,1]$ and treatment assignment is established by the exposure cut-off $\pi$.  
At each visit $k$, subjects in poor health with $P_i \ge \pi$ are deterministically assigned to exposure and are considered positivity non-compliant. In contrast, subjects in poor health with $P_i < \pi$ have a positive probability of either receiving exposure or remaining unexposed and are considered positivity compliant.
This leads to near-violations of the positivity assumption because, within the poor health subgroup defined by individuals $i$ with $L_{i,k}\in \mathcal{I}_{\tau}$, both exposure and unexposure are theoretically possible. However, non-exposure may be rarely observed due to the randomness in drawing the individual propensities for exposure, creating situations where non-compliance with positivity occurs more frequently than expected, especially for low values of $\pi$.

The value $\pi$ indeed represents the expected proportion of subjects for whom both exposure and non-exposure are observable in all subgroups defined by the measured confounder. In other words, it can be interpreted as the expected proportion of subjects compliant with positivity, and therefore:
\begin{itemize}
	\item $\pi=1$ means that  all subjects are compliant with positivity, representing no positivity violations;
	\item $0 < \pi < 1$ means that   $\pi\times 100\%$ subjects are expected to be compliant with positivity, representing near-positivity violations;
	\item $\pi=0$ means that  all subjects are non-compliant with positivity, representing strict-positivity violations.
\end{itemize}
Therefore, for $\mathcal{I}_{\tau}$ fixed, the higher the cut-off $\pi$, the less severe the violation.

Given an algorithm to simulate longitudinal survival data from MSMs in presence of time-varying confounding, near-positivity violations can be incorporated by using the pseudocode structure in panel Algorithm \ref{alg:posviol}. The main advantage of imposing near-positivity violations in an existing approach, where longitudinal data generation from the desired hazard-MSM has already been confirmed, is the ability to directly examine the impact on IPTW estimators solely attributable to positivity violations, rather than other sources of bias. 
In this way, the original data-generating mechanism can be considered as the ``truth'' or benchmark case where the positivity assumption holds.

\begin{algorithm}
	\caption{General pseudocode for positivity violation.\label{alg:posviol}}
	\begin{algorithmic}
		\State Initialize parameters: $(\mathcal{I}_{\tau},\pi,\dots)$
		\For{$i=1,\dots,n$}
		\State $\dots$
		\State $P_i \sim \mathcal{U}(0,1)$ \Comment{Draw the individual propensity}
		\For{$k=0,\dots,K$}
		\State $L_{i,k}$ is assigned based on the generating algorithm
		\If{$P_i \ge \pi$ and $L_{i,k}\in \mathcal{I}_{\tau}$} 
		\State exposure is assigned deterministically: $A_{i,k}$ = 1
		\Else
		\State exposure $A_{i,k}$ is assigned stochastically based on the generating algorithm
		\EndIf
		\State $\dots$
		\EndFor
		\EndFor
	\end{algorithmic}
\end{algorithm}

\section{Simulation study I}\label{sec:simuI}

\subsection{Data generation}\label{sec:simuI:alg}
The first algorithm proposed in this work is based on the data-generating mechanism introduced by \cite{Havercroft2012} to simulate from a discrete-time logit-MSM. This mechanism is now briefly introduced and then extended by imposing positivity violations.

\subsubsection{Benchmark I in a nutshell}\label{sec:simuI:alg:Benchmark}
Building upon the DAG in Figure \ref{fig:dags}a, \cite{Havercroft2012} proposed an algorithm  to emulate longitudinal data from the Swiss HIV Cohort Study \citep{Sterne2005}. The authors considered a discrete-time setting where visit times correspond to visit numbers, i.e., $q_k=k$ for all $k=0,\dots,K$.
The time-dependent binary treatment process $A_{i,k}$ represents exposure to the highly active antiretroviral therapy (HAART) versus no treatment (unexposure). Once HAART has started for a subject $i$, it continues until failure or end of the follow-up period.  The only measured time-dependent confounder $L_{i,k}$ is the non-negative CD4 cell count, measured in $cells/\mu L$.  Variable $U_{i,k}$ represents the individual general latent health process, indicating a poor individual health status at visit $k$ for values close to 0, or good health conditions for values close to 1. The latent process $U_{i,k}$ and the survival process $Y_{i,k+1}\in \{0,1\}$ are updated at each time point $k$, whereas CD4 cell count $L_{i,k}\ge 0$ and HAART exposure $A_{i,k} \in \{0,1\}$ are updated every $\kappa$-th time points, for a chosen $\kappa$, named check-up visits. 

Despite there being no direct arrow from ${L}_{i,k}$ to ${Y}_{i,k+1}$, the DAG (Figure \ref{fig:dags}a) exhibits time-dependent confounding due to $U_{i,0}$ being a common ancestor of $\bar{\boldsymbol{A}}_{i}$ via $\bar{\boldsymbol{L}}_{i}$ and $\bar{\boldsymbol{Y}}_{i}$. Moreover, $A_{i,k}$ is independent from $\bar{\boldsymbol{U}}_{i,k}$ given $\left(\bar{\boldsymbol{L}}_{i,k},\bar{\boldsymbol{A}}_{i,k-1}\right)$ and the vector $\bar{\boldsymbol{L}}_{i}$ is sufficient to adjust for confounding. Based on this mechanism, the authors proposed an algorithm to correctly simulate data from the following discrete-time logit-MSM:
\begin{align} \label{eq:logitMSM:algI}
	\lambda_{k}^{\bar{\boldsymbol{a}}} &= \text{logit}^{-1}\left[\tilde\gamma_{0}+\tilde\gamma_{A1} \cdot \{(1-a_{k})k + a_{k}k^{*}\}+\tilde\gamma_{A2} \cdot a_{k} + \tilde\gamma_{A3} \cdot a_{k}(k-k^{*})\right]  \nonumber \\ 
	&= \text{logit}^{-1}\left[\tilde\gamma_{0}+\tilde\gamma_{A1} \cdot d_{1k}+\tilde\gamma_{A2} \cdot a_{k} + \tilde\gamma_{A3} \cdot d_{3k}\right],
\end{align}
where $a_k$ is the binary treatment strategy at time $k$, $k^{*}$ is the treatment initiation time, $d_{1k} = \min\{k,k^{*}\}$ and $d_{3k} = \max\{k-k^{*},0\}$ represent the time elapsed before and after treatment initiation, respectively. Note that $g\left(\tilde{\boldsymbol{\gamma}}_A;
\bar{\boldsymbol{a}}_{k}\right) = \tilde\gamma_{A1}\cdot d_{1k}+\tilde\gamma_{A2}\cdot a_{k} + \tilde\gamma_{A3}\cdot d_{3k}$ in \eqref{eq:logitMSM:algI}, reflecting that the hazard-MSM depends on a summary of the treatment history rather than only on the current treatment.
In particular, Havercroft and Didelez proved that the parameters $(\tilde\gamma_{0},\tilde\gamma_{A1},\tilde\gamma_{A2},\tilde\gamma_{A3})$ in the desired logit-MSM \eqref{eq:logitMSM:algI} are collapsible with the conditional distribution parameters $(\gamma_{0},\gamma_{A1},\gamma_{A2},\gamma_{A3})$ in the following conditional logit model:
\begin{equation}\label{eq:condlogit:algI}
	\lambda_{i,k} = \text{logit}^{-1}\left[\gamma_{0}+\gamma_{A1}\cdot \left\{(1-A_{i,k})k + A_{i,k}K_{i}^{*}\right\}+\gamma_{A2}\cdot A_{i,k} + \gamma_{A3} \cdot A_{i,k}(k-K_{i}^{*})\right],
\end{equation}
where $K_i^{*}$ is the individual treatment initiation time, and $\lambda_{i,k}$ represents the individual probability of failure in the interval $k<t\le k+1$ conditional on survival up to visit $k$.

\subsubsection{Algorithm I: imposing random positivity violations in Benchmark I}\label{sec:simuI:alg:viol}

As illustrated in the DAG in Figure \ref{fig:dags}c, the first proposed algorithm to account for potential near-positivity violations builds upon Benchmark I (Figure \ref{fig:dags}a) by incorporating two additional components.
\begin{itemize}
	\item[(i)] First, a poor health subgroup identified by $\mathcal{I}_{\tau}$ and acting on the purple path $L_{i,k} \rightarrow A_{i,k}$ must be defined.
	Since a CD4 count below 500 $cells/\mu L$ indicates the patients' immune system may be weakened, making them susceptible to developing serious infections from viruses, bacteria, or fungi that typically do not cause problems in healthy individuals, it is reasonable to assume that subjects in a poor health condition at visit $k$ are identified by $L_{i,k} < \tau$, where $\tau \in [0;500] \, cells/\mu L$. This is equivalent to a non-negative CD4 range of the form $\mathcal{I}_{\tau} = [0,\tau)$, where the upper threshold $\tau$ has to be defined according to the simulation scenario. The higher the upper threshold $\tau$, the wider $\mathcal{I}_{\tau}$ and the more severe the violations.
	
	\item[(ii)] Then, the latent individual propensity for exposure $P_{i}\sim \mathcal{U}(0,1)$ directly acts on $A_{i,k}$. Subjects in poor health condition with propensity $P_i$ above the \textit{exposure cut-off} $\pi$ (to be defined according to the simulation scenario) are forced to start the treatment.
	
\end{itemize}

The procedure proposed below extends the algorithm by \cite{Havercroft2012} by incorporating the possibility of near-positivity violations. For details regarding the chosen parameter values, please refer to their primary work.
\begin{description}
	\item[\textit{Procedure}] 
	For each subject $i=1,\dots,n$, the simulation procedure with $K$ discrete time points and check-ups every $\kappa$-th visits is as follows.
	
	\begin{enumerate}
		\item Generate the individual propensity to exposure: $P_{i} \sim \mathcal{U}(0,1)$.
		\item Generate the general latent health status at baseline: $U_{i,0} \sim \mathcal{U}(0,1)$.
		\item Generate the baseline CD4 as a transformation of $U_{i,0}$ by the inverse cumulative distribution function of $\Gamma(3,154)$ distribution plus an error $\epsilon_{i,0}\sim \mathcal{N}(0,20)$: $L_{i,0} = F^{-1}_{\Gamma(3,154)}(U_{i,0}) + \epsilon_{i,0}$.
		\item If $P_i \ge \pi$ and $L_{i,0}< \tau$, the subject starts HAART and $A_{i,0} = 1$. Otherwise, draw treatment decision $A_{i,0} \sim Be \left(p^A_{i,0}\right)$ where $p^A_{i,0} = logit^{-1} \left[ -0.405 -0.00405 \cdot (L_{i,0}-500)\right]$. If $A_{i,0} = 1$, set the treatment initiation time $K^{*}_{i}$ to $0$.
		\item Compute the conditional individual hazard $\lambda_{i,0}$ for $k=0$ using \eqref{eq:condlogit:algI}. If $\lambda_{i,0} \geq U_{i,0}$, death has occurred in the interval $(0,1]$ and set $Y_{i,1} = 1$. Otherwise, the subject survived and set $Y_{i,1} = 0$.
	\end{enumerate}
	If the individual is still at risk at visit $k=1$:
	\begin{enumerate}
		\setcounter{enumi}{5}
		\item Draw $U_{i,k} = \min\{1,\max\{0,U_{i,k-1}+\mathcal{N}(0,0.05)\}\}$ as a perturbation of $U_{i,k-1}$ restricted to $[0,1]$.
		\item If $k$ is not a check-up visit, CD4 cell count are not updated and $L_{i,k} = L_{i,k-1}$. Otherwise, update the count as $L_{i,k} = \max\left\{0,L_{i,k-1}+150\cdot A_{i,k-1}+\epsilon_{i,k}\right\}$, where the addition of 150 indicates the positive effect of exposure to HAART on CD4 count, and $\epsilon_{i,k}\sim\mathcal{N}(100(U_{i,k}-2),50)$ is a Gaussian drift term implying that the worse is the general health condition $U_{i,k}$ (i.e., value closer to 0), the stronger the negative drift in CD4.
		\item If $k$ is not a check-up visit, treatment is not updated and $A_{i,k} = A_{i,k-1}$. Otherwise, assign exposure
		\begin{itemize}
			\item[a.] \textit{deterministically}: if $P_i \ge \pi$ and $L_{i,k}< \tau$ or if treatment has started at previous check-up ($A_{i,k-\kappa}=1$), patient $i$ is exposed to HAART and $A_{i,k} = 1$;
			\item[b.] \textit{stochastically}: otherwise, draw treatment decision $A_{i,k} \sim Be \left(p^A_{i,k}\right)$, where $$p^A_{i,k} = logit^{-1} \left[-0.405 + 0.0205 \cdot k -0.00405 \cdot (L_{i,k}-500)\right].$$ The conditional distribution parameters has been set to calibrate the logistic function such that $\Pr(A_{\bullet,0} = 1\mid L_{\bullet,0} = 500) = 0.4$, $\Pr(A_{\bullet,0} = 1\mid L_{\bullet,0} = 400) = 0.5$, and $\Pr(A_{\bullet,10} = 1\mid L_{\bullet,10} = 500) = 0.45$.
		\end{itemize}
		If the subject starts the treatment at visit/time $k$, set the treatment initiation time $K^{*}_{i}$ equal to $k$.
		\item Compute $\lambda_{i,k}$, i.e., the individual probability of failure in the interval $(k; k+1]$ conditional on survival up to visit $k$, using \eqref{eq:condlogit:algI}.
		If $S_i(t)=\prod_{j = 0}^{k}(1-\lambda_{i,j}) \leq 1-U_{i,0}$, the death has occurred in the interval  $(k; k+1]$ and $Y_{i,k+1} = 1$. Otherwise, the subject remains at risk and $Y_{i,k+1} = 0$.
		\item Repeat steps 6–9 for $k=2,\dots,K$.
	\end{enumerate}
\end{description}

The related pseudocode is provided in Appendix A.1. Note that when $\pi=1$ the positivity assumption always holds and  this procedure corresponds to the data generating mechanism of Benchmark I.

\subsection{Simulation study using Algorithm I}\label{sec:simuI:res}
\subsubsection{Methods and estimands}\label{sec:simuI:res:meth}
Investigations are performed in several scenarios by considering different sample sizes ($n=$ 50, 100, 250, 500, 1000), exposure cut-off values ($\pi=$ 0.05, 0.1, 0.3, 0.5, 0.8, 1), WT strategies (NoWT, 1-99, 5-95, 10-90), and  poor health subgroups $\mathcal{I}_{\tau} = [0;\tau)$  with varying upper thresholds ($\tau =$ 0, 100, 200, 300, 400, 500 measured in $cells/\mu L$).
The other parameters are set to be identical to those used by \cite{Havercroft2012} to consider their results as a benchmark for this analysis. 
Specifically, $K=40$ time-points with check-ups every ($\kappa=5$)-th visit are considered, and the desired conditional distribution parameters in Equation \eqref{eq:condlogit:algI} are $(\gamma_{0},\gamma_{A1},\gamma_{A2},\gamma_{A3}) = (-3,0.05,-1.5,0.1)$. In this way, the true values of the parameters in logit-MSM \eqref{eq:logitMSM:algI} are $(\tilde\gamma_{0}^*,\tilde\gamma_{A1}^*,\tilde\gamma_{A2}^*,\tilde\gamma_{A3}^*) = (-3,0.05,-1.5,0.1)$.

For each scenario, $B=1000$ simulated datasets are generated. The logit-MSM \eqref{eq:logitMSM:algI} is fitted to each simulated dataset through IPTW estimation using (truncated) stabilized weights. Weight components at time $k$ are estimated by logistic regression models for the probability of treatment initiation at time $k$, with numerator and denominators in \eqref{eq:sw} defined respectively as:
\begin{align*}
	\Pr\left(A_{i,k} = 1 \mid \boldsymbol{\bar{A}}_{i,k-1}= \boldsymbol{0}, Y_{i,k-1}=0\right) &= logit^{-1}\left[\theta_{0} + \theta_{1}\cdot k\right] \quad \text{ and }\\
	\Pr\left(A_{i,k} = 1\mid \boldsymbol{\bar{A}}_{i,k-1}= \boldsymbol{0},\boldsymbol{\bar{L}}_{i,k}, Y_{i,k-1}=0\right) &= logit^{-1}\left[\theta_{0} + \theta_{1}\cdot k + \theta_{2}\cdot(L_{i,k}-500)\right].
\end{align*} 
In this way, since $\pi \neq 0$, the denominator model is correctly specified according the data-generating mechanism \citep{Bryan2004,Havercroft2012}. 

The estimands of interest are the regression coefficients $\left(\tilde{\gamma}_0,\tilde{\boldsymbol{\gamma}}_A\right)$ and the marginal survival probabilities in Equation \eqref{eq:MSMlogit:surv} for the \textit{always treated} versus \textit{never treated} regimens, where $g\left(\tilde{\boldsymbol{\gamma}}_A;
\bar{\boldsymbol{a}}_{k}\right) = \tilde\gamma_{A1}\cdot d_{1k}+\tilde\gamma_{A2}\cdot a_{k} + \tilde\gamma_{A3}\cdot d_{3k}$.
For each regression coefficient, estimated bias, empirical Standard Error (empSE), and Root Mean Squared Error (RMSE) \citep{Morris2019} are considered as performance measures.
Marginal survival curves are presented graphically by showing the mean estimated curves across the $B=1000$ repetitions.

Note that simulation settings with $\pi=1$ and NoWT are equivalent to Benchmark I, regardless of $\tau$ (positivity always holds when $\pi=1$). In such cases, the analyses are based on correctly specified logit-MSMs and correctly specified models for the weights, so the resulting estimates are expected to be approximately unbiased.

\subsubsection{Results}\label{sec:simuI:res:results}
The estimated performance metrics (bias, empSE, RMSE) for the regression coefficients $\left(\hat{\tilde{\gamma}}_{0},\hat{\tilde{\gamma}}_{A1},\hat{\tilde{\gamma}}_{A2},\hat{\tilde{\gamma}}_{A3}\right)$ are provided in Supplementary Material S1.
Smaller sample sizes exhibit worse performance. While increasing the sample size reduces bias and variability caused by finite sample limitations, estimation errors resulting from violations still persist, particularly for low values of $\pi$. Across all scenarios, as the severity of violations increases (i.e., the bigger $\tau$ and the lower $\pi$), the absolute bias, empSE, and RMSE grow substantially, particularly for for $\tilde\gamma_{0}$ and $\tilde\gamma_{A2}$ (i.e., the intercept and the parameter most directly related to the effect of exposure). This outcome reflects how large weights resulting from near-violations increase variability and reduce precision in IPTW-based estimates. Adopting a WT strategy reduces variability and slightly decreases bias by truncating extreme weights, especially for larger $\tau$. However, further narrowing the truncation range (e.g., from WT 1–99 to WT 5–95 or 10–90) does not lead to additional improvements in performance; indeed, it may lead to poorer performance.

\begin{figure}[h!]
	\centering
	\includegraphics[width=0.95\textwidth]{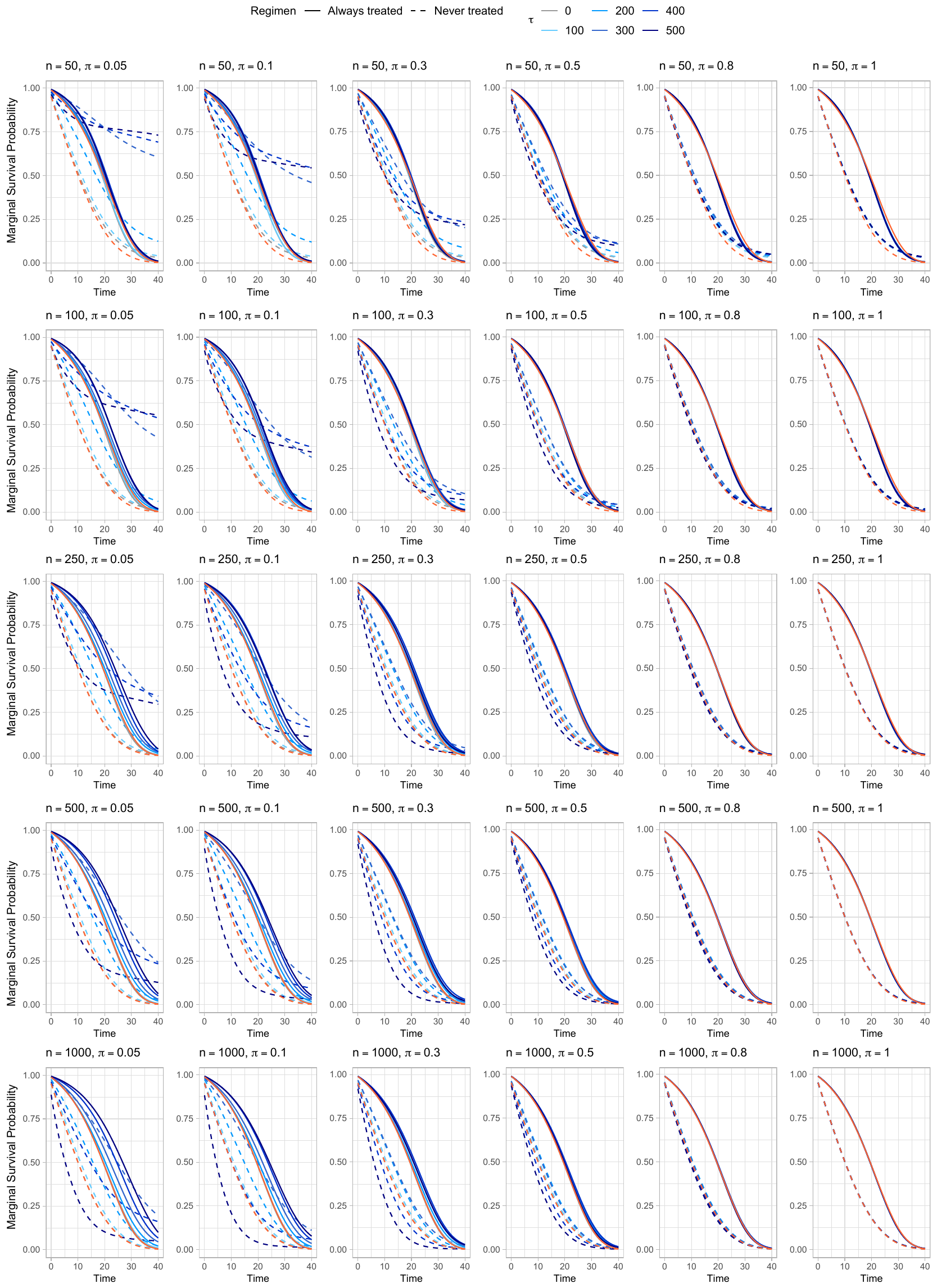}
	\caption{Marginal survival probability curves averaged across all the $B=1000$ repetitions for different settings without weight truncation (No WT) of simulation study I. Each row refers to a different sample size $n=50,100,250,500,1000$. Each column refers to a different exposure cut-off $\pi=0.05,0.1,0.3,0.5,0.8,1$. Dashed lines refer to the \textit{never treated} regimen, while solid ones to the \textit{always treated} regimen. Curves are coloured according to different values of rule-threshold $\tau$. True marginal survival curves are shown in orange.}  \label{fig:aI:surv:n}
\end{figure}

\begin{figure}[h!]
	\centering
	\includegraphics[width=0.95\textwidth]{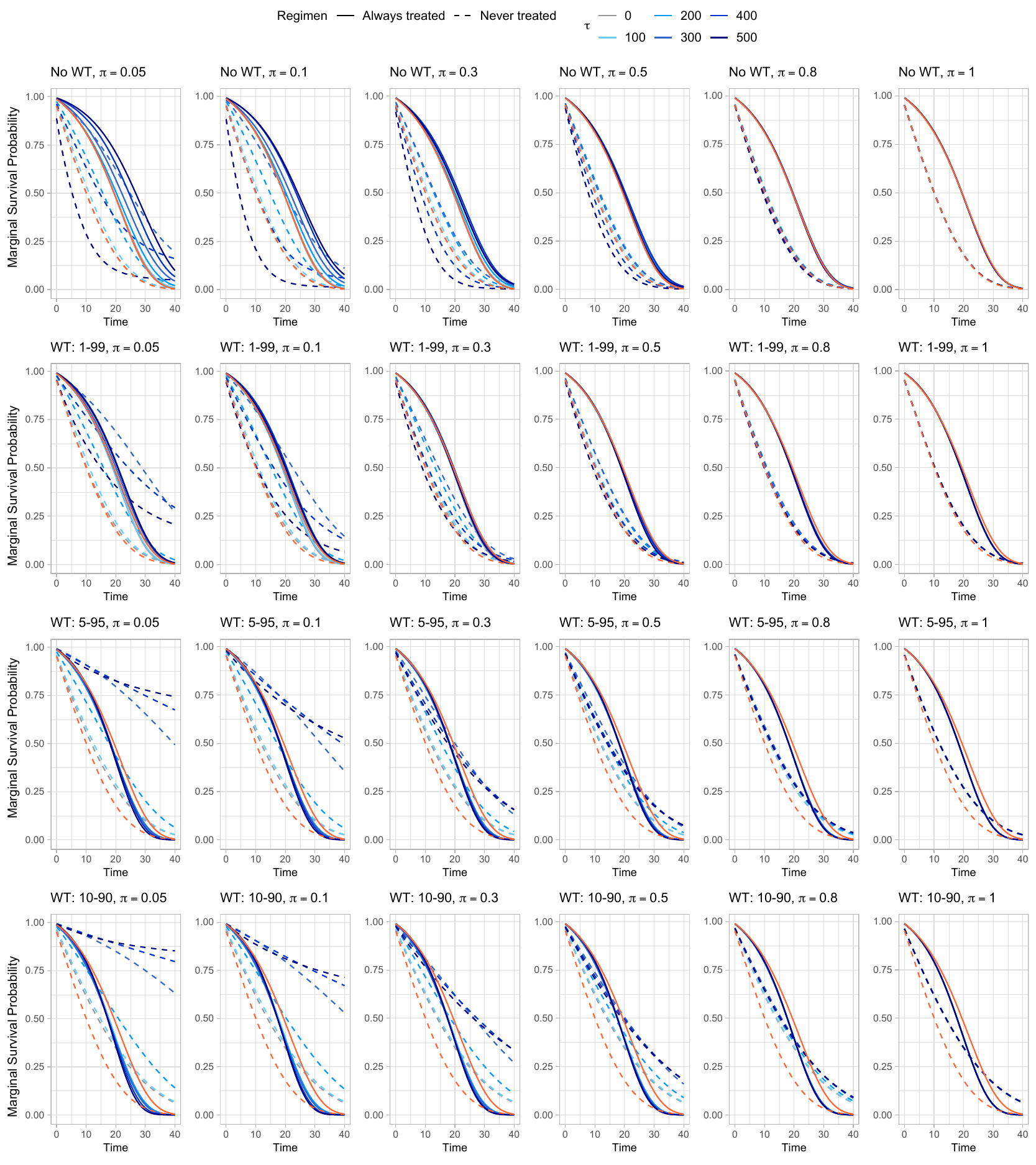}
	\caption{Marginal survival probability curves averaged across all the $B=1000$ repetitions for different settings with sample size $n=1000$ of simulation study I. Each row refers to a different weight truncation (WT) strategy: No WT, 1-99, 5-95, 10-99. Each column refers to a different exposure cut-off $\pi=0.05,0.1,0.3,0.5,0.8,1$. Dashed lines refer to the \textit{never treated} regimen, while solid ones to the \textit{always treated} regimen. Curves are coloured according to different values of rule-threshold $\tau$. True marginal survival curves are shown in orange.}  \label{fig:aI:surv:wt}
\end{figure}

Figures \ref{fig:aI:surv:n} shows the mean marginal survival curves, along with the true ones (in orange), in each simulated scenario without WT. 
Each line refers to a different $\tau$ value; the darker the line colour, the more severe the violation (i.e., the bigger $\tau$).  Each row refers to a different sample size ($n=50,100,250,500,1000$) and each column refers to a different exposure cut-off ($\pi=0.05,0.1,0.3,0.5,0.8,1$).  The magnitude and direction of the bias in each regression coefficient (see Supplementary Figure S1) determine how closely the estimated marginal survival curves match the true (orange) ones (see Equation \eqref{eq:MSMlogit}).
Curves for \textit{never treated} (dashed lines) depend on $(\hat{\tilde{\gamma}}_{0},\hat{\tilde{\gamma}}_{A1})$ and therefore generally show a greater deviation compared to those of \textit{always treated} (solid lines) which depend on $(\hat{\tilde{\gamma}}_{0},\hat{\tilde{\gamma}}_{A2},\hat{\tilde{\gamma}}_{A3})$.
Indeed, in the \textit{always treated} the negative bias of $\hat{\tilde{\gamma}}_{0}$ can be balanced by the positive one of $\hat{\tilde{\gamma}}_{A2}$. This is not true for the \textit{never treated} so the deviation notably increases as near-positivity violations become more concrete (i.e., as $\pi$ decreases). 
Similar behaviours are observed across different sample sizes as $\pi$ varies. All scenarios eventually converged to the true values when no positivity violations are present.
Contrary to what should happen, for more severe violations ($\pi=0.05,0.1$; $\tau=300,400,500$), the \textit{never treated} group demonstrates notably better survival curves compared to the \textit{always treated} group. This deviation becomes even more pronounced with small sample sizes or when any WT strategy is applied, as illustrated in Figure  \ref{fig:aI:surv:wt}, where each row corresponds to a different WT strategy with a sample size of $n=1000$ .  The estimated mean curves are very close to the true ones for an expected positivity compliance rate of 80\% under NoWT, or even for 50\% under 1-99 WT. However, under 5-95 or 10-95 WT, estimated curves deviate from true ones, even for a high exposure cut-off.

\section{Simulation study II}\label{sec:simuII}

\subsection{Data generation}\label{sec:simuII:alg}
The second algorithm proposed in this work is based on the data-generating mechanism introduced by \cite{Keogh2021} to simulate from an Aalen-MSM. This mechanism is now briefly introduced and then extended by imposing positivity violations.

\subsubsection{Benchmark II in a nutshell}\label{sec:simuII:alg:Benchmark}
\cite{Keogh2021} proposed a setting where, at each visit $k=0,\dots,K$, a binary treatment process $A_{i,k} \in \{0,1\}$ (control vs treatment) and a time-dependent biomarker $L_{i,k}\in \mathbb{R}$ are observed for each subject $i$. The assumed data structure is illustrated in Figure \ref{fig:dags}b using a discrete-time DAG setting where visit times correspond to visit numbers (i.e., $q_k=k$ $\forall k$) and $Y_{i,k+1} = I(k< T_i \le k+1)$ is an indicator of whether the event $T_i$ occurs between visits $k$ and $k+1$. As the time intervals become very small the algorithm approaches the continuous time setting.
The DAG also includes a baseline latent variable $U_i$, representing a subject-specific unmeasured individual frailty, which has a direct effect on $L_{i,k}$ and $Y_{i,k+1}$, but not on $A_{i,k}$.

The DAG in Figure \ref{fig:dags}b exhibits time-dependent confounding due to $L_{i,k}$ that predicts subsequent treatment use $A_{i,k}$, is affected by earlier treatment $A_{i,k-1}$, and affects the outcome $Y_{i,k+1}$ through pathways that are not just through subsequent treatment. Moreover, because $U_i$ is not a confounder of the association between the treatment and the outcome, the fact that it is unmeasured does not affect the ability to estimate causal effects of treatments.
The authors demonstrated that using a conditional additive hazard of the form
\begin{equation}\label{eq:condaalen:algII}
	\lambda_{i}\left(t \mid \bar{\boldsymbol{A}}_{i,\lfloor t \rfloor},\bar{\boldsymbol{L}}_{i,\lfloor t \rfloor}, U_{i}\right) = \alpha_0 + \alpha_A \cdot A_{i,\lfloor t \rfloor} + \alpha_L \cdot L_{i,\lfloor t \rfloor}  + \alpha_U \cdot U_i,
\end{equation}
their data generating mechanism correctly simulates data from the additive Aalen-MSM of the form:
\begin{equation}\label{eq:MSMaalen:algII}
	\lambda^{\bar{\boldsymbol{a}}}(t) = \tilde{\alpha}_{0}(t)+ \sum_{j=0}^{{\lfloor{t}\rfloor}} \tilde{\alpha}_{Aj}(t) \cdot a_{{\lfloor{t}\rfloor}-j},
\end{equation}
that is an Aalen-MSM including as treatment-pattern $g\left(\tilde{\boldsymbol{\alpha}}_A(t);\bar{\boldsymbol{a}}_{\lfloor{t}\rfloor}\right)$ the main effect terms at each visit (see Table \ref{tab:gform}). 
Researchers using this approach can only specify the parameters $(\alpha_{0},\alpha_{A},\alpha_{L},\alpha_{U})$ of the conditional model \eqref{eq:condaalen:algII}. The true values of the cumulative regression coefficients $C_0(t) = \int_{0}^{t}\tilde{\alpha}_{0}(s)ds$ and $C_{Aj}(t)=\int_{0}^{t}\tilde{\alpha}_{Aj}(s)ds$ ($j=0,\dots,4$) of the Aalen-MSM \eqref{eq:MSMaalen:algII} must be computed using a simulation-based approach, as detailed in \cite{Keogh2021}.

Thanks to the collapsibility property of the Aalen's additive hazard model, the generating mechanism of Benchmark II includes the direct arrow from $L_{i,k}$ to $Y_{i,k+1}$, making it more realistic in practice compared to Benchmark I.
Unlike Benchmark I, which is restricted to generating data closely matching the Swiss HIV Cohort Study \citep{Sterne2005}, Benchmark II can hence be applicable in more general contexts. However, its parameter values need to be carefully selected to ensure that the chance of obtaining a negative hazard, a common drawback in Aalen's model, is negligible.

\subsubsection{Algorithm II: imposing random positivity violations in Benchmark II}\label{sec:simuII:alg:viol}
Analogously to Algorithm I, the second algorithm extends Benchmark II (Section \ref{sec:simuII:alg:Benchmark}) by imposing positivity violations randomly. As illustrated in the DAG in Figure \ref{fig:dags}d, compared to Benchmark II's structure in Figure \ref{fig:dags}b, two components are introduced.
\begin{itemize}
	\item[(i)] The latent individual propensity for exposure $P_{i}\sim \mathcal{U}(0,1)$ directly acts on $A_{i,k}$: subjects in poor health condition with propensity $P_i$ above the \textit{exposure cut-off} $\pi$ (to be defined according to the simulation scenario) are always exposed.
	
	\item[(ii)] The poor health subgroup identified by $\mathcal{I}_{\tau}$ acts on the purple path $L_{i,k} \rightarrow A_{i,k}$. In the framework of Keogh et al.'s procedure, the confounder $L_{i,k}$ represents a general biomarker
	with no direct real-world interpretation, yet its higher values at time $k$ correspond to an increased likelihood of exposure and a increased hazard.	It is hence reasonable to assume that subjects with with a poor health condition at time $k$ are identified by $L_{i,k} > \tau$. This is equivalent to a range $\mathcal{I}_{\tau} = (\tau,\infty)$, where the lower threshold $\tau$ has to be defined according to the simulation scenario. Here, the lower the threshold $\tau$, the wider the interval $\mathcal{I}_{\tau}$ and the more severe the violation.
\end{itemize}

The proposed procedure extends Keogh et al.'s algorithm by incorporating the possibility for near-positivity violations. For details regarding the chosen parameter values, please refer to their primary work.
\begin{description}
	\item[\textit{Procedure}]  For each subject $i=1,\dots,n$, the simulation procedure with $K+1$ as administrative censoring time is as follows.
	\begin{enumerate}
		
		\item Generate the individual propensity to exposure: $P_{i} \sim \mathcal{U}(0,1)$.
		\item Generate the individual frailty term: $U_{i} \sim \mathcal{N}(0,0.1)$.
		\item Generate the baseline biomarker as a transformation of $U_{i}$: $L_{i,0} \sim \mathcal{N}(U_{i},1)$.
		\item If $P_i \ge \pi$ and $L_{i,0}> \tau$, the subject is exposed to treatment and $A_{i,0} = 1$. Otherwise, draw treatment decision $A_{i,0} \sim Be \left(p^A_{i,0}\right)$ where $p^A_{i,0} = logit^{-1} \left[-2 +0.5 \cdot L_{i,0}\right]$. 
		\item Event times in the period $0<t<1$ are generated by calculating $\Delta_{i} = -\log(\upsilon_{i,0})/\lambda_{i}\left(t \mid A_{i,0},L_{i,0},U_{i}\right)$, where at numerator $\upsilon_{i,0} \sim \mathcal{U}(0,1)$ and the denominator is the individual conditional hazard in \eqref{eq:condaalen:algII} with $\lfloor t \rfloor=0$ and desired parameters. If $\Delta_{i} <1$, death occurred in the interval $t \in (0,1)$: the event time is set to be $T_{i} = \Delta_{i}$  and the failure process is $Y_{i,1} = 1$. Otherwise, subjects with  $\Delta_{i} \ge 1$ remain at risk at time $t=1$ and set $Y_{i,1} = 0$.
	\end{enumerate}
	If the individual is still at risk at visit $k=1$:
	\begin{enumerate}
		\setcounter{enumi}{5}
		\item Update the biomarker value as: $L_{i,k} \sim \mathcal{N}(0.8\cdot L_{i,k-1} - A_{i,k-1} +0.1 \cdot k + U_{i}, 1)$.
		\item Assign exposure
		\begin{enumerate}
			\item[a.] \textit{deterministically}: if $P_i \ge \pi$ and $L_{i,k}> \tau$, subject $i$ is exposed to treatment and $A_{i,k} = 1$;
			\item[b.] \textit{stochastically}: otherwise, draw treatment decision $A_{i,k} \sim Be \left(p^A_{i,k}\right)$ where $$p^A_{i,k} = logit^{-1} \left[-2 + 0.5 \cdot L_{i,k} + A_{i,k}\right].$$
		\end{enumerate}
		
		\item Event times in the period $k\le t<k+1$ are generated by calculating $\Delta_{i} = -\frac{\log(\upsilon_{i,k})}{\lambda_{i}\left(t \mid \bar{\boldsymbol{A}}_{i,k},\bar{\boldsymbol{L}}_{i,k},U_{i}\right)}$, where $\upsilon_{i,k} \sim \mathcal{U}(0,1)$ and the denominator is the individual conditional hazard in \eqref{eq:condaalen:algII} with $\lfloor t \rfloor=k$ and desired parameters. If $\Delta_{i} <1$, death occurred in the interval $[k,k+1)$: the event time is set to be $T_{i} = k+\Delta_{i}$  and the failure process is $Y_{i,k+1} = 1$. Otherwise, subjects with  $\Delta_{i} \ge 1$ remain at risk at time $k+1$, i.e., $Y_{i,k+1} = 0$.
		
		\item Repeat steps 6–8 for $k=2,\dots,K$. Subjects who do not have an event time generated in the period $0<t<K+1$ are administratively censored at time $K+1$.
		
	\end{enumerate}    
\end{description} 

The related pseudocode is provided in Appendix A.2.  Note that when $\pi=1$ the positivity assumption always holds and this procedure corresponds to the data generating mechanism of Benchmark II.

\subsection{Simulation study using Algorithm II}\label{sec:simuII:res}

\subsubsection{Methods and estimands}\label{sec:simuII:res:meth}

Investigations are performed in several scenarios by considering different sample sizes ($n=$ 50, 100, 250, 500, 1000), exposure cut-off values ($\pi=$ 0, 0.05, 0.1, 0.3, 0.5, 0.8, 1), WT strategies (NoWT, 1-99, 5-95, 10-90), and poor health subgroups identified by intervals $\mathcal{I}_{\tau} = (\tau; \infty)$ with varying lower threshold $\tau$. Since $L_{i,k}$ represents a general biomarker with no direct real-world interpretation,  the choice of possible values for $\tau$ relies on the distribution of the complete history of biomarker values generated using Benchmark II with 100,000 subjects. Specifically, the rounded values closest to the 80th, 90th, 95th, 99th, and 100th percentiles (i.e., $\tau=1,1.5,2,3,7$) plus an extreme value outside the observed range (i.e, $\tau=10$) are considered as possible lower thresholds.
The other parameters are set to be identical to those considered by \cite{Keogh2021} in order to (i) have the same true values of the estimands of interest for the Aalen-MSM \eqref{eq:MSMaalen:algII} (see Tables 1 and 2 in \cite{Keogh2021}), (ii) use their results as a benchmark for this analysis, and (iii) ensure that the probability of obtaining a negative hazard is negligible for Benchmark II.
Specifically, $K=4$ time-points with administrative censoring at $K+1$ are considered, and the conditional distribution parameters in Equation \eqref{eq:condaalen:algII} were $(\alpha_{0},\alpha_{A},\alpha_{L},\alpha_{U}) = (0.7,-0.2,0.05,0.05)$.

For each scenario, $B=1000$ simulated datasets are generated. The Aalen-MSM \eqref{eq:MSMaalen:algII} is fitted to each simulated dataset through IPTW estimation using (truncated) stabilized weights. Weight components at time $k$ are estimated by logistic regression models for the probability of being exposed at time $k$, with numerator and denominators in \eqref{eq:sw} defined respectively as:
\begin{align*}
	\Pr\left(A_{i,k} = 1 \mid \boldsymbol{\bar{A}}_{i,k-1}, T_i \ge k\right) &= logit^{-1}\left[\theta_{0} + \theta_{1}\cdot A_{i,k-1}\right] \quad \text{ and }\\ 
	\Pr\left(A_{i,k} = 1\mid \boldsymbol{\bar{A}}_{i,k-1},\boldsymbol{\bar{L}}_{i,k}, T_{i}\ge k\right) &= logit^{-1}\left[\theta_{0} + \theta_{1}\cdot A_{i,k-1} + \theta_{2}\cdot L_{i,k}\right].
\end{align*}
In this way, since $\pi \neq 0$, the denominator model is correctly specified according the data generation mechanism \citep{Keogh2021}.

The estimands of interest are the cumulative regression coefficients $C_0(t) = \int_{0}^{t}\tilde{\alpha}_{0}(s)ds$ and $C_{Aj}(t)=\int_{0}^{t}\tilde{\alpha}_{Aj}(s)ds$ ($j=0,\dots,4$), and the marginal survival probabilities in Equation \eqref{eq:MSMaalen:surv} for the \textit{always treated} and \textit{never treated} regimens, where $g\left(\tilde{\boldsymbol{\alpha}}_A(t);\bar{\boldsymbol{a}}_{\lfloor{t}\rfloor}\right) = \sum_{j=0}^{{\lfloor{t}\rfloor}} \tilde{\alpha}_{Aj}(t) \cdot a_{{\lfloor{t}\rfloor}-j}$.
For the cumulative regression coefficients, results are presented graphically by showing the performance (i.e., bias, empSE, and RMSE) measured at times $t=1,2,3,4,5$. 
For the marginal survival curves, the mean value of the estimates across repetitions are presented graphically across time points $t=1,2,3,4,5$.

Note that simulation settings with $\pi=1$ and NoWT are equivalent to Benchmark II, regardless of $\tau$ (positivity always holds). In such cases, the analyses are based on correctly specified Aalen-MSMs and correctly specified models for the weights, so the resulting estimates are expected to be approximately unbiased.

\subsubsection{Results}\label{sec:simuII:res:results}

The performance metrics for the estimated cumulative coefficients over time $t=1,2,3,4,5$ in each simulated scenario are provided in Supplementary Material S2.
For cumulative coefficient $\hat{C}_{A0}(t)$ related to the current main effect terms, smaller sample sizes exhibit worse performance, as increasing the sample size only mitigates the bias induced by finite sample issues. 
Across scenarios, bias, empSE, and RMSE increase with time. In general, the more severe the violation (i.e., low $\pi$, low $\tau$), the higher empSE and RMSE. Adopting a WT strategy decreases empSE and RMSE, as extreme weights are truncated, especially for more severe violations. However, compared to WT 1-99, narrowing down the WT resulted in worse bias over time. Results eventually converge to be unbiased under NoWT, but a small bias still persists  under 1-99 WT for $t=4,5$. In terms of variability, the empSE are comparable to the one estimated in Benchmark II, even when the expected positivity non-compliance rate is 80\%.
Results for the other cumulative coefficients led to similar conclusions: (i) the performance worsen with time; (ii) adopting a WT strategy reduces the variability; (ii) estimated performance eventually converge to the ones estimated for the Benchmark II. 
The estimated bias for $\hat{C}_{0}(t)$, i.e., the cumulative intercept, is negative, and decreases with time (see Supplementary Figure S4). For large sample sizes ($n=500,1000$), the difference in the bias across $\pi$ values is minimal. The other treatment-related coefficients $\hat{C}_{Aj}(t)$ with $j=1,2,3,4$ exhibit higher empSE and RMSE as the violation severity increases. Unlike $\hat{C}_{A0}(t)$, however, no discernible relationship was found between the values of $\tau$ and the resulting bias.

\begin{figure}[h!]
	\centering
	\includegraphics[width=0.95\textwidth]{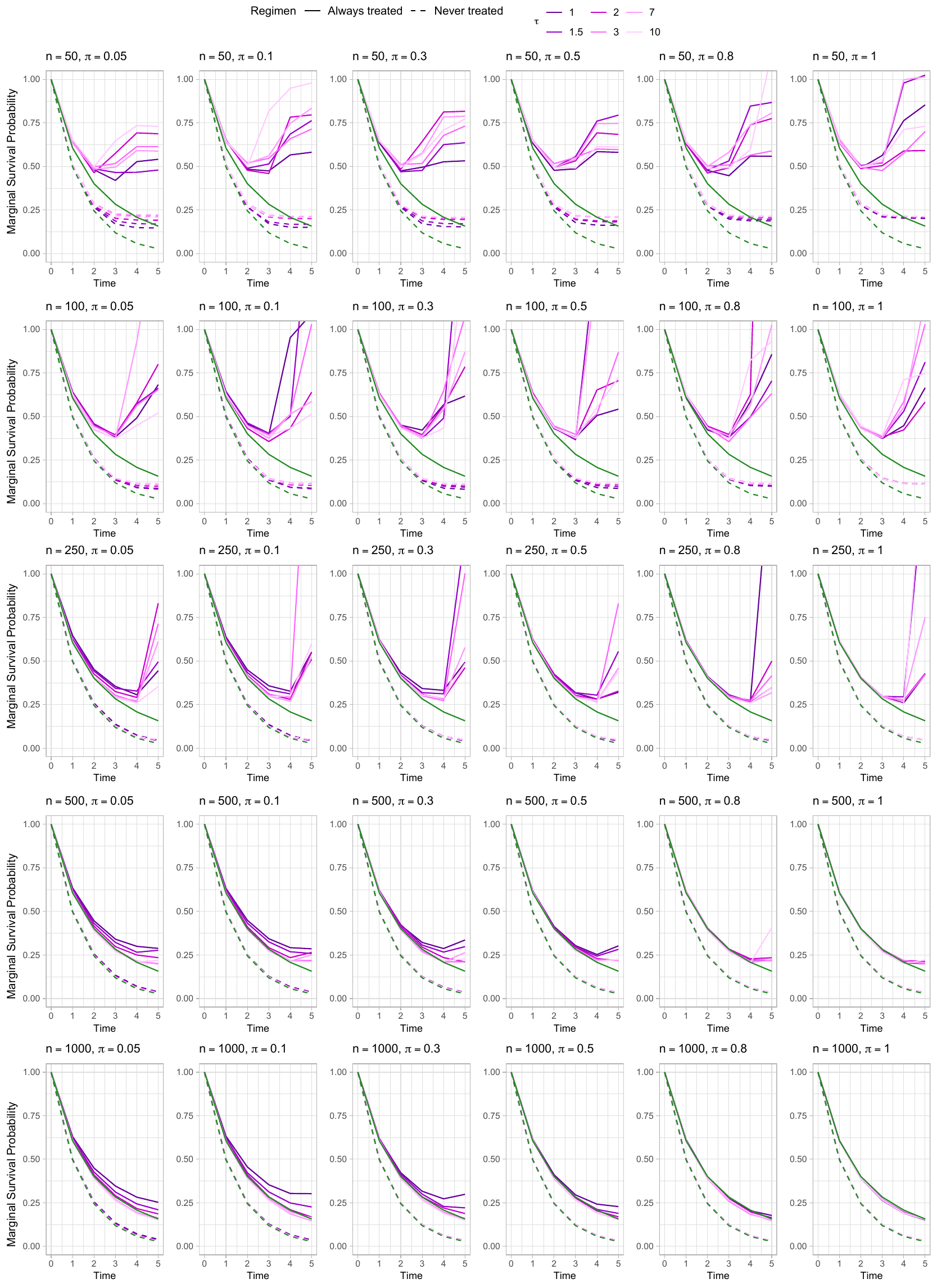}
	\caption{Marginal survival probability curves averaged across all the $B=1000$ repetitions for different settings without weight truncation (No WT) of simulation study II. Each row refers to a different sample size $n=50,100,250,500,1000$. Each column refers to a different exposure cut-off $\pi=0.05,0.1,0.3,0.5,0.8,1$. Dashed lines refer to the \textit{never treated} regimen, while solid ones to the \textit{always treated} regimen. Curves are coloured according to different values of rule-threshold $\tau$. True marginal survival curves are shown in green.}  \label{fig:aII:surv:n}
\end{figure}

\begin{figure}[h!]
	\centering
	\includegraphics[width=0.95\textwidth]{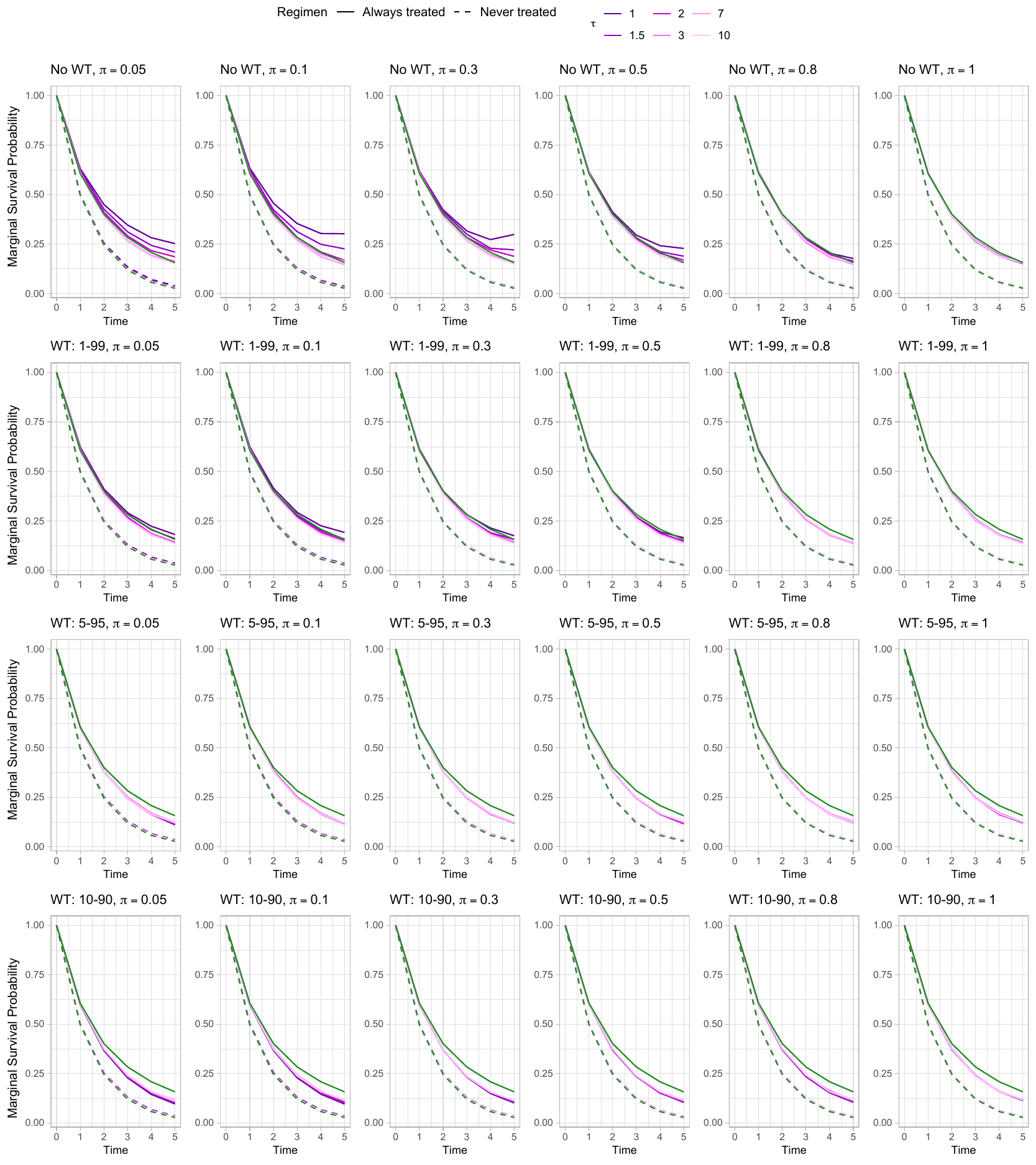}
	\caption{Marginal survival probability curves averaged across all the $B=1000$ repetitions for different settings with sample size $n=1000$ of simulation study II. Each row refers to a different weight truncation (WT) strategy: No WT, 1-99, 5-95, 10-99. Each column refers to a different exposure cut-off $\pi=0.05,0.1,0.3,0.5,0.8,1$. Dashed lines refer to the \textit{never treated} regimen, while solid ones to the \textit{always treated} regimen. Curves are coloured according to different values of rule-threshold $\tau$. True marginal survival curves are shown in green.}  \label{fig:aII:surv:wt}
\end{figure}

Figure \ref{fig:aII:surv:n} shows the mean marginal survival curves over times $t=1,2,3,4,5$ across repetitions estimated in the various scenarios without WT, along with the true ones (in green). 
Each line refers to a different $\tau$ value; the darker the line colour, the more severe the violation (i.e., the smaller $\tau$).  Each row refers to a different sample size ($n=50,100,250,500,1000$) and each column to a different exposure cut-off ($\pi=0.05,0.1,0.3,0.5,0.8,1$).  These curves can be derived from the cumulative coefficients -- as in Equation \eqref{eq:MSMaalen:surv} -- whose estimates determine how closely the estimated mean curves match the true (green) ones.
At each time point $t=1,2,3,4,5$, the curves for \textit{never treated} (dashed lines) depend solely on $\hat{C}_{0}(t)$, while all cumulative coefficients contribute to estimating the curves for \textit{always treated} (solid lines). As a result, the estimated mean curves for \textit{never treated} align with the true (green) curves for big sample sizes, exhibiting very low bias across time points. Conversely, summing the contributions of each cumulative coefficients leads to higher bias for the \textit{always treated}, especially at later time points. Small sample sizes heavily suffer from the main drawback of the additive hazard model, which does not restrict the hazard to be non-negative. This determines survival probabilities for the \textit{always treated} that wrongly increase over time. This issue is mitigated for bigger samples sizes ($n=500,1000$). As the exposure cut-off $\pi$ increases, the estimated mean curves correspond to the true ones for an expected positivity compliance rate from 80\% upwards  under NoWT, or even from 30\% under 1-99 WT.
Adopting 1-99 WT (see Figure \ref{fig:aII:surv:wt}) improves the performance compared to NoWT. Nonetheless, compared to WT 1-99, narrowing down the WT resulted in worse bias over time. 

\section{Discussion}\label{sec:discussion}
Simulation studies play a key role in evaluating robustness to assumption violations, enabling the examination of various properties \citep{Morris2019,Friedrich2023}. 
While existing literature on positivity violations in MSMs has largely focused on incorrect inferences using real data or simulations with exposure assigned at a single or two time points, this study fills the gap by presenting two simulation studies in realistic survival contexts involving a time-varying binary treatment and a continuous time-dependent confounder. Two distinct algorithms were proposed to simulate data from hazard-MSMs and to account for potential near-positivity violations, where remaining unexposed is rare within certain confounder levels.
Systematic simulations were conducted to evaluate the impact of near-positivity violations on the performance of target estimands obtained via IPTW under various scenarios and WT strategies.

Findings from both studies revealed a consistent trend: as the violation becomes more severe (i.e., low $\pi$), performance deteriorates. Increasing the sample size mitigates bias and variability due to finite sample size, but incorrect inference resulting from positivity violations persists. Even when $\mathcal{I}_{\tau}$ is small, performance may still be poor due to the presence of extreme weights. Under NoWT, the higher the positivity compliance rate, the better the performance aligns with Benchmark I and II.
Adopting a WT strategy always reduces variability by truncating extreme weights, especially for wider $\mathcal{I}_{\tau}$. However, compared to WT 1-99, narrowing down the WT may not improve bias. This suggests that bias becomes the more dominant factor when the positivity assumption is violated.
The decision to adopt the 1-99 WT strategy in cases of near-violations should carefully consider the bias-variance trade-off. 
For intermediate positivity compliance rates ($\pi = 0.3, 0.5$), the 1-99 WT strategy generally outperformed NoWT. In contrast, for high values ($\pi = 0.8, 1$), NoWT was more effective.

Algorithm I and II proposed in this work were built on prior algorithms by \cite{Havercroft2012} and \cite{Keogh2021}, respectively.
The advantage of extending existing algorithms was threefold. First, the issue of non-collapsibility \citep{Robinson1991,Didelez2022} between conditional and marginal models and the replication of complex confounding dynamics have been already overcome in the original studies. Second, by controlling the exposure-confounder path and avoiding misspecification of the weighting model, the effect due to the imposed positivity violations was separated from other sources of bias. Third, the original Benchmarks I and II were used as the references for the expected true estimates when positivity is valid (i.e., for $\pi=1$).

While a direct comparison is not feasible as they pertain to different data-generating mechanisms, Algorithm I generally exhibited poorer performance compared to Algorithm II. This difference may stem from their distinct treatment decision mechanisms. Algorithm I requires continuous exposure until failure or censoring once treatment begins, whereas Algorithm II does not have such a requirement. This constraint limits the possible combinations of treatment-covariate history in Algorithm I, with a significant impact on the estimated coefficients even though very few combinations are missing. Consequently, this influences the estimated mean survival curves, leading to incorrect survival probabilities for the \textit{never treated} group. 
On the other hand, Algorithm II suffers, especially with small sample sizes, from the linear form of the Aalen-MSM, which does not restrict the hazard to be non-negative, resulting in unrealistic survival estimates.

This work has its limitations, which also open up intriguing possibilities for future research. 
Both studies focused on instances where violations occur within a single interval of the confounder variable and examined only a single continuous confounding variable. However, in real-world scenarios, violations may span varied intervals, and multiple continuous/categorical confounding factors are typically present. This highlights interesting directions for extending the proposed algorithms, though adapting them to new contexts will require meticulous adjustments.
Nonetheless, in their current form, Algorithms I and II developed in this study represent a valuable contribution to the literature. They could serve as data-generating tools for systematic analyses, enabling for (i) the comparison of different techniques for estimating causal effects from observational data under near-positivity violations and (ii) the evaluation of potential new methods designed to address near-positivity violations in a longitudinal-treatment framework. Since current methods for detecting and addressing positivity violations are primarily tailored to point-treatment settings \citep{Traskin2011,Karavani2019,Zhu2023,Danelian2023,Zivich2024}, developing methodologies specifically suited to a longitudinal framework presents a challenging direction for future research.

In summary, this study emphasizes the importance of carefully assessing positivity compliance to ensure robust and reliable causal inference in survival studies, while also highlighting the risks of underestimating it. 
By demonstrating the substantial impact of near-positivity violations, it underscores the need for rigour in causal inference, particularly given the exponential growth of causal inference approaches and their applications to observational data. In practical analyses, researchers are strongly encouraged to examine group-wise descriptives for the original and weighted populations, utilize bootstrap to quantify uncertainty in weights, and conduct sensitivity analysis for further insights \citep{colehernan2008,Austin2015,Desai2019}.
The causal effect of interest must be defined with consideration of positivity violations. While adopting a WT strategy may reduce variability, it should be approached with caution due to the potential risk of increased bias. 
Although IPTW-based MSMs are widely used in applied studies for their simplicity in implementation and interpretation, analysts must remain vigilant about blindly accepting the positivity assumption, as doing so can lead to detrimental consequences. Finally, the two algorithms developed in this study also serve as valuable tools for generating data in future systematic analyses of novel causal inference methodologies.

\vspace{1cm}
\small
\noindent\textbf{Acknowledgments.}
The author gratefully acknowledges  Prof.dr. Marta Fiocco (Mathematical Institute, Leiden University) for her valuable initial inputs and discussions on this research.


\newpage 
\appendix
\section*{Appendix}

\subsection*{A.1\enspace Pseudocode of Algorithm I}\label{app:algI}
\setcounter{algorithm}{0}
\renewcommand\thealgorithm{\Roman{algorithm}}

\begin{algorithm}\small
	\caption{\enskip Pseudocode of Algorithm I introduced in Section \ref{sec:simuI:alg:viol}.}\label{alg:simI}
	\begin{algorithmic}
		\State \textbf{Input parameters:} 
		\State ~~$n=$ sample size, $\tau=$ rule-threshold, $\pi=$ compliance threshold
		\State ~~$K= $ number of visits, $\kappa = $ check-up times, 
		\State ~~$(\gamma_{0},\gamma_{A1},\gamma_{A2},\gamma_{A3}) =$ conditional distribution parameters in Equation  \eqref{eq:condlogit:algI} 
		\hrule
		\State \textbf{Algorithm:} 
		\For {each subject $i=1,\dots, n$}
		\State $P_{i} \sim \mathcal{U}(0,1)$ \Comment{Individual propensity}
		\State $U_{i,0} \sim \mathcal{U}(0,1)$
		\State $L_{i,0} = F^{-1}_{\Gamma(3,154)}(U_{i,0}) + \epsilon_{i,0}$ where $\epsilon_{i,0}\sim \mathcal{N}(0,20)$
		\If{$P_i \ge \pi$ and $L_{i,0}< \tau$}
		\State $A_{i,0} = 1$ \Comment{Deterministic exposure assignment}
		\Else
		\State $p^A_{i,0} = logit^{-1}\left[-0.405 -0.00405\cdot(L_{i,0}-500)\right]$ \Comment{Stochastic exposure assignment}
		\State $A_{i,0} \sim Be\left(p^A_{i,0}\right)$ 
		\EndIf  
		\If {$A_{i,0} = 1$} $K^{*}_{i} = 0$ \EndIf
		\State $\lambda_{i,0}=logit^{-1}\left[\gamma_{A0} +\gamma_{A2}\cdot A_{i,0}\right]$  \Comment{Conditional hazard \eqref{eq:condlogit:algI}}
		\If {$\lambda_{i,0} \ge U_{i,0}$} $Y_{i,1}=1$ \Else $Y_{i,1}=0$ \EndIf
		\State $k=1$
		\While{$Y_{i,k}=0$ and $k \le K$}
		\State  $U_{i,k} = \min\{1,\max\{0,U_{i,k-1}+\varepsilon_{i,k}\}\}$ where $\varepsilon_{i,k} \sim \mathcal{N}(0,0.05)$
		\If{$k$ mod $\kappa \neq 0$}
		\State $L_{i,k} = L_{i,k-1}$ 
		\State $A_{i,k} = A_{i,k-1}$ 
		\Else
		\State $L_{i,k} = \max\left\{0,L_{i,k-1}+150\cdot A_{i,k-1}+\epsilon_{i,k}\right\}$ where $\epsilon_{i,k}\sim\mathcal{N}(100(U_{i,k}-2),50)$
		\If{($P_i \ge \pi$ and $L_{i,k}< \tau$) or $A_{i,k-\kappa}=1$}
		\State $A_{i,k}=1$ \Comment{Deterministic exposure assignment}
		\Else
		\State $p^A_{i,k} = logit^{-1}\left[-0.405 +0.0205 \cdot k -0.00405 \cdot (L_{i,k}-500)\right]$ \Comment{Stochastic exposure assignment}
		\State $A_{i,k} \sim Be\left(p^A_{i,k}\right)$
		\EndIf
		\If {$A_{i,k} = 1$ and $A_{i,k-1} = 0$} $K^{*}_{i} = k$ \EndIf
		\EndIf
		\Comment{Conditional hazard \eqref{eq:condlogit:algI}}
		\State $\lambda_{i,k} = logit^{-1}\left[\gamma_{0}+\gamma_{A1} \cdot \left\{(1-A_{i,k})k + A_{i,k}K_{i}^{*}\right\}+\gamma_{A2} \cdot A_{i,k} + \gamma_{A3} \cdot A_{i,k}(k-K_{i}^{*})\right]$ 
		\If {$\prod_{j = 0}^{k}(1-\lambda_{i,j})\leq 1 - U_{i,0}$} $Y_{i,k+1}=1$ \Else $Y_{i,k+1}=0$ \EndIf
		\State $k=k+1$
		\EndWhile
		\EndFor
	\end{algorithmic}
\end{algorithm}

\newpage
\subsection*{A.2\enspace Pseudocode of Algorithm II}\label{app:algiI}
\begin{algorithm} \small
	\caption{\enskip Pseudocode of Algorithm II introduced in Section \ref{sec:simuII:alg:viol}.}\label{alg:simII}
	\begin{algorithmic}
		\State \textbf{Input parameters:} 
		\State ~~$n=$ sample size, $\tau=$ rule-threshold, $\pi=$ compliance threshold,
		\State ~~$K= $ number of visits, $(\alpha_{0},\alpha_{A},\alpha_{L},\alpha_{U})  =$ desired true parameters in Equation \ref{eq:condaalen:algII}
				\hrule
		\State \textbf{Algorithm:} 
		\For {each subject $i=1,\dots, n$}
		\State $P_{i} \sim \mathcal{U}(0,1)$ \Comment{Individual propensity}
		\State $U_{i,0} \sim \mathcal{N}(0,1)$
		\State $L_{i,0} \sim \mathcal{N}(U_i,1)$
		\If{$P_i \ge \pi$ and $L_{i,0}> \tau$}
		\State $A_{i,0} = 1$ \Comment{Deterministic exposure assignment}
		\Else
		\State $p^A_{i,0} = logit^{-1} \left[-2 + 0.5 \cdot L_{i,0} \right]$ \Comment{Stochastic exposure assignment}
		\State $A_{i,0} \sim Be\left(p^A_{i,0}\right)$ 
		\EndIf  
		\State $\lambda_{i}\left(t \mid A_{i,0}, L_{i,0}, U_{i}\right) = \alpha_0 + \alpha_A \cdot A_{i,0} + \alpha_L \cdot L_{i,0}  + \alpha_U \cdot U_i$  \Comment{Conditional hazard \eqref{eq:condaalen:algII}}
		
		\State $\Delta_{i} = -\log(\upsilon_{i,0})/\lambda_{i}(t \mid A_{i,0},L_{i,0},U_{i})$ where $\upsilon_{i,0} \sim \mathcal{U}(0,1)$
		\If {$\Delta_{i} <1$} $T_{i} = \Delta_{i}$ and $Y_{i,1}=1$ \Else $Y_{i,1}=0$ \EndIf
		\State $k=1$
		\While{$Y_{i,k}=0$ and $k \le K$}
		\State $L_{i,k} \sim \mathcal{N}(0.8\cdot L_{i,k-1} - A_{i,k-1} +0.1 \cdot k + U_{i}, 1)$
		\If{$P_i \ge \pi$ and $L_{i,k}> \tau$}
		\State $A_{i,k}=1$ \Comment{Deterministic exposure assignment}
		\Else
		\State $p^A_{i,k} = logit^{-1} \left[-2 + 0.5 \cdot L_{i,k} + A_{i,k}\right]$ \Comment{Stochastic exposure assignment}
		\State $A_{i,k} \sim Be\left(p^A_{i,k}\right)$
		\EndIf
		
		\State $\lambda_{i}\left(t \mid \bar{\boldsymbol{A}}_{i,k},\bar{\boldsymbol{L}}_{i,k},U_{i}\right) = \alpha_0 + \alpha_A \cdot A_{i,k} + \alpha_L \cdot L_{i,k}  + \alpha_U \cdot U_i$ \Comment{Conditional hazard \eqref{eq:condaalen:algII}}
		
		\State $\Delta_{i} = -\log(\upsilon_{i,k})/\lambda_{i}(t \mid A_{i,0},L_{i,0},U_{i})$ where $\upsilon_{i,k} \sim \mathcal{U}(0,1)$
		\If {$\Delta_{i} <1$} $T_{i} = k+ \Delta_{i}$ and $Y_{i,k+1}=1$ \Else $Y_{i,k+1}=0$ \EndIf
		
		\State $k=k+1$
		\EndWhile
		\If {is.null($T_i$)} $T_{i} = K+1$ \EndIf \Comment{Administrative censoring at $K+1$}
		\EndFor
	\end{algorithmic}
\end{algorithm}

\newpage
\setcounter{page}{1} 
\setcounter{figure}{0} 
\begin{center}
	\LARGE \textbf{Supplementary Material}
\end{center}
\renewcommand{\thefigure}{S\arabic{figure}}
\renewcommand{\thesection}{S\arabic{section}}
\renewcommand{\thepage}{S\arabic{page}}
\section{Simulation study I: additional results}
\begin{figure}[!h]
	\includegraphics[width=0.86\textwidth]{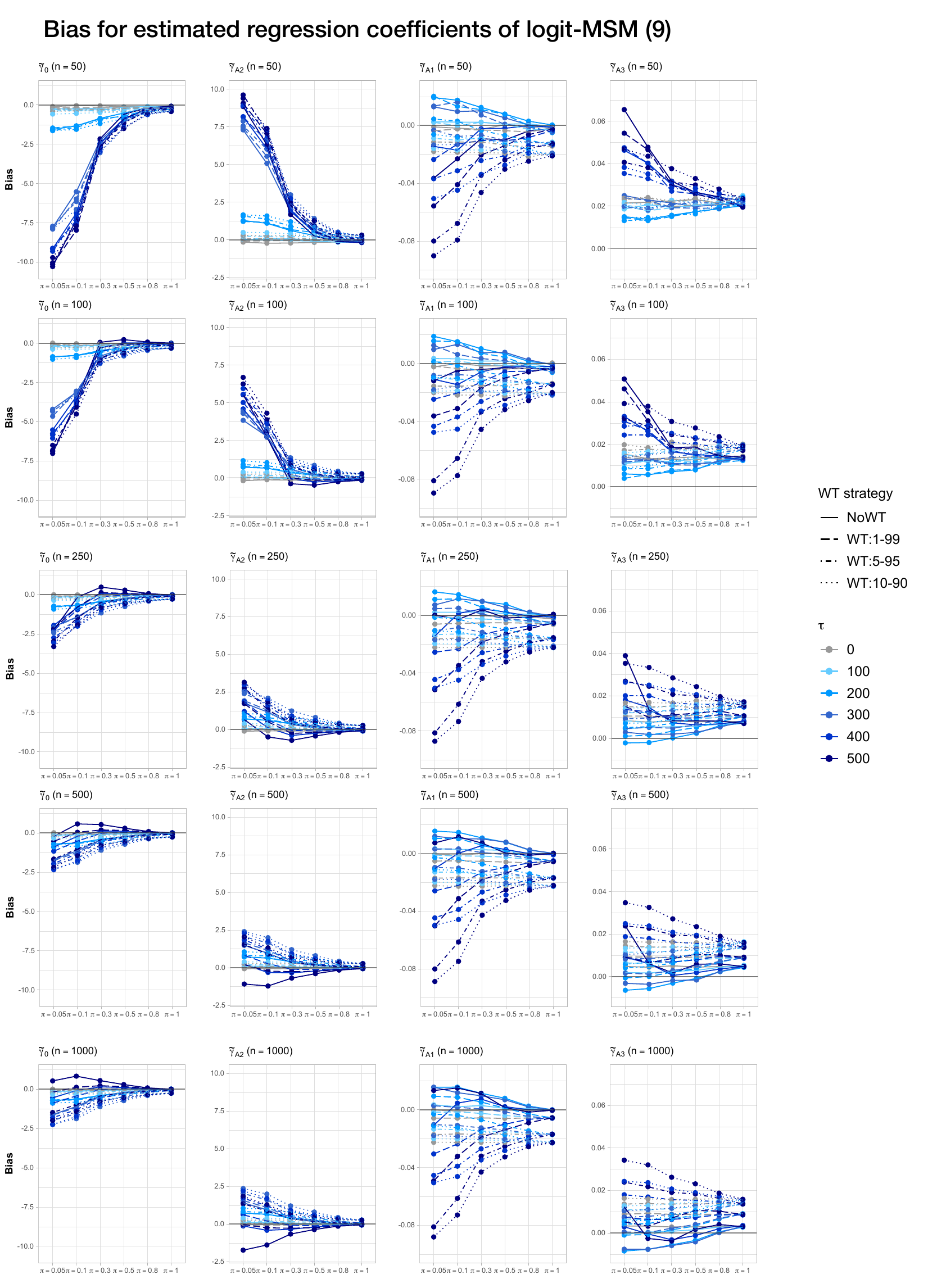}
	\caption{Bias of the coefficient estimates for the different setting of simulation study I. Each column refer to a different coefficient ($\tilde\gamma_{0}$: first column;  $\tilde\gamma_{A1}$: third column; $\tilde\gamma_{A2}$: second column; $\tilde\gamma_{A3}$: fourth column). Each row refers to a different sample size $n=50,100,250,500,1000$.  The x-axes show the compliance-threshold values $\pi$. Different types of line refer to different weight truncation (WT) strategies (solid: No WT; long-dashed: 1-99 WT; dot-dashed: 5-95 WT; dotted: 10-90 WT). The colours refer to different values of the rule-threshold $\tau$: the darker the colour, the more severe the violation (i.e., the higher $\tau$). Note that the ranges of y-axes differ between panels.} \label{fig:coefI:bias}
\end{figure}

\newpage
\begin{figure}[!h]
	\includegraphics[width=0.86\textwidth]{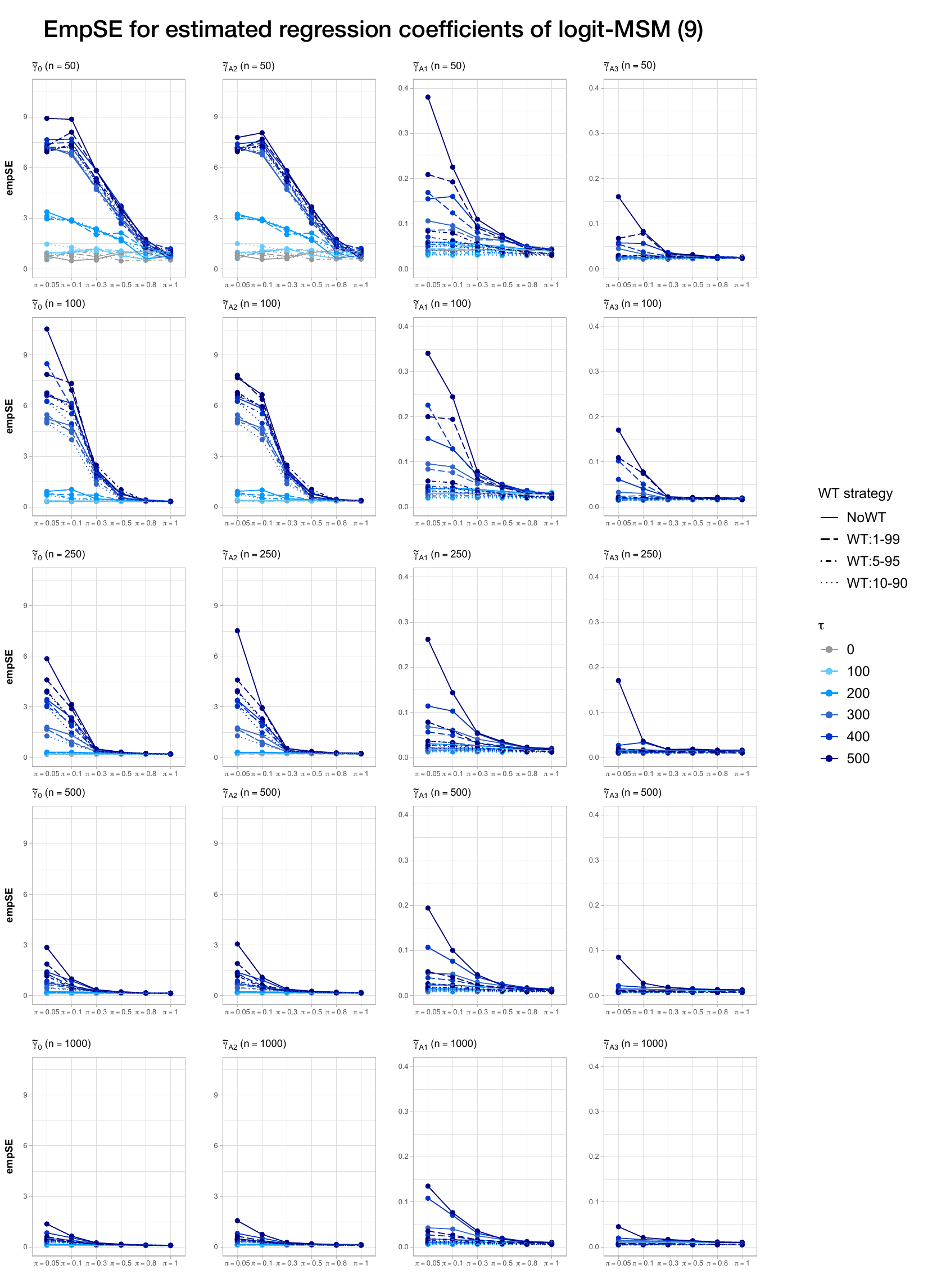}
	\caption{Empirical Standard Error (empSE) of the coefficient estimates for the different setting of simulation study I. Each column refer to a different coefficient ($\tilde\gamma_{0}$: first column;  $\tilde\gamma_{A1}$: third column; $\tilde\gamma_{A2}$: second column; $\tilde\gamma_{A3}$: fourth column). Each row refers to a different sample size $n=50,100,250,500,1000$.  The x-axes show the compliance-threshold values $\pi$. Different types of line refer to different weight truncation (WT) strategies (solid: No WT; long-dashed: 1-99 WT; dot-dashed: 5-95 WT; dotted: 10-90 WT). The colours refer to different values of the rule-threshold $\tau$: the darker the colour, the more severe the violation (i.e., the higher $\tau$). Note that the ranges of y-axes differ between panels.} \label{fig:coefI:empse}
\end{figure}

\newpage
\begin{figure}[!h]
	\includegraphics[width=0.86\textwidth]{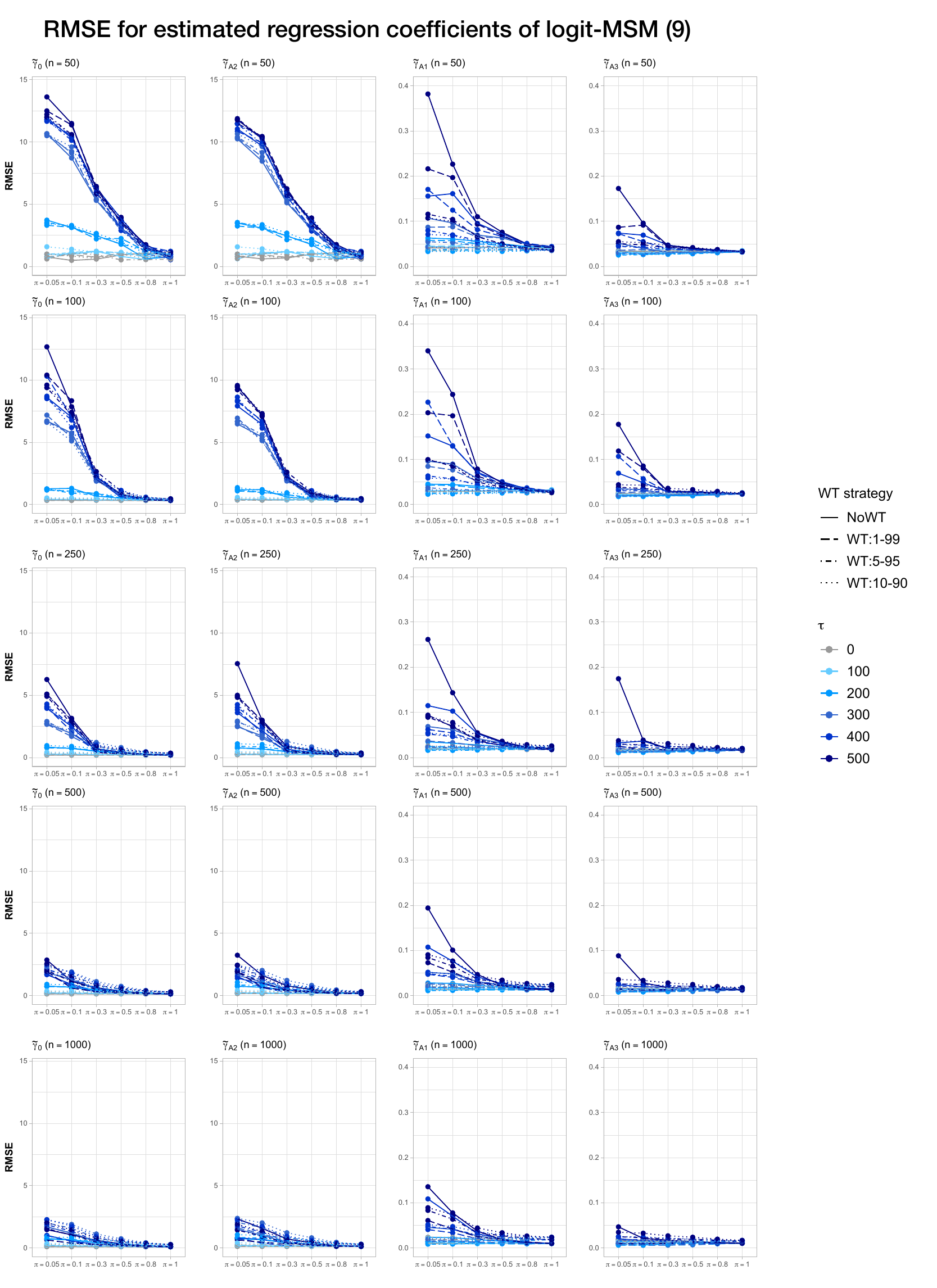}
	\caption{Root Mean Squared Error (RMSE) of the coefficient estimates for the different setting of simulation study I. Each column refer to a different coefficient ($\tilde\gamma_{0}$: first column;  $\tilde\gamma_{A1}$: third column; $\tilde\gamma_{A2}$: second column; $\tilde\gamma_{A3}$: fourth column). Each row refers to a different sample size $n=50,100,250,500,1000$.  The x-axes show the compliance-threshold values $\pi$. Different types of line refer to different weight truncation (WT) strategies (solid: No WT; long-dashed: 1-99 WT; dot-dashed: 5-95 WT; dotted: 10-90 WT). The colours refer to different values of the rule-threshold $\tau$: the darker the colour, the more severe the violation (i.e., the higher $\tau$). Note that the ranges of y-axes differ between panels.} \label{fig:coefI:rmse}
\end{figure}

\newpage
\section{Simulation study II: additional results}
\subsection{Bias}
\begin{figure}[!h]
	\includegraphics[width=0.9\textwidth]{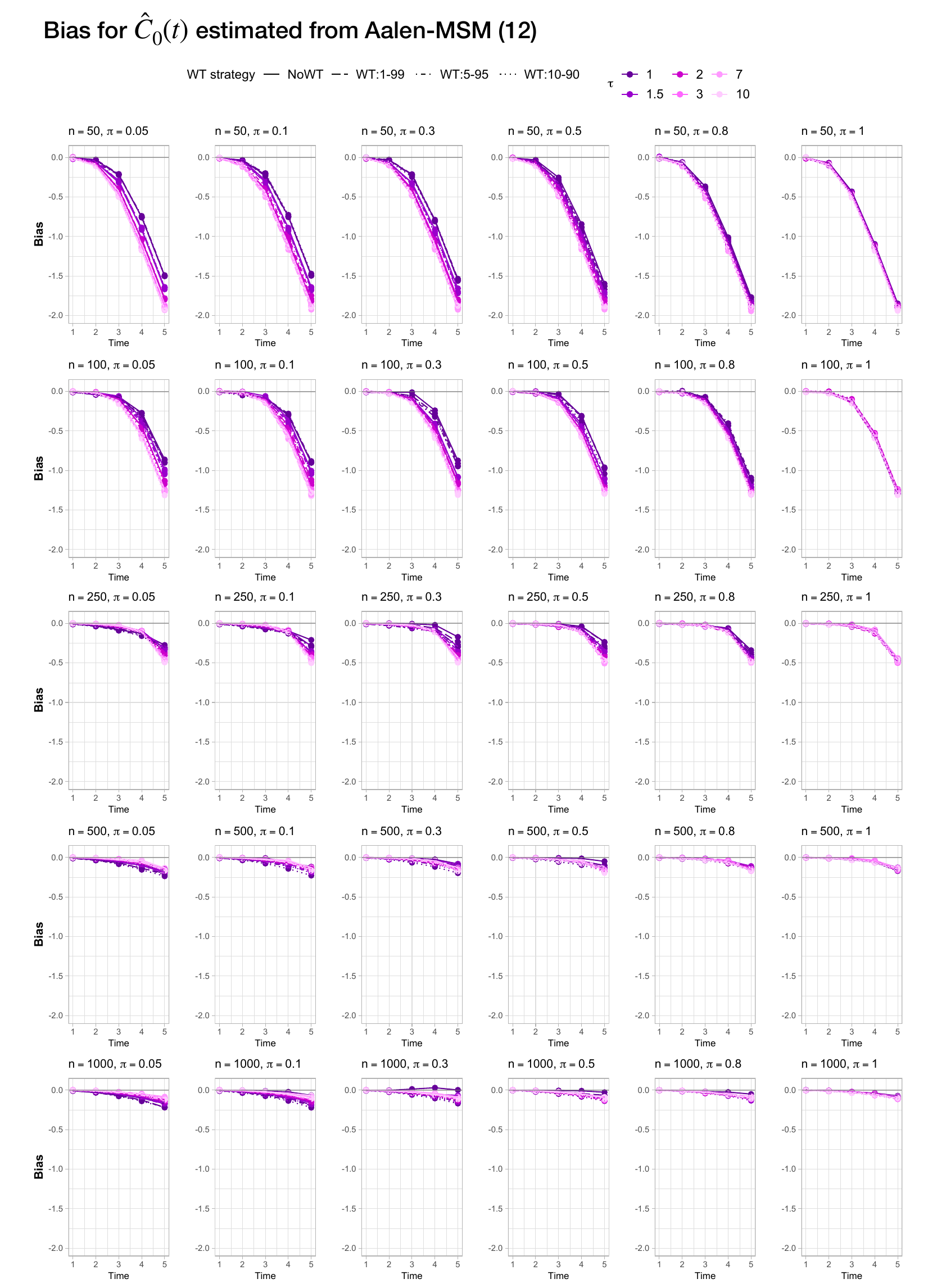}
	\caption{Bias of the estimates for the cumulative coefficient $C_{0}(t) =\int_{0}^{t}\tilde{\alpha}_{0}(s)ds$ at time points $t=1,\dots,5$ for the different setting of simulation study II. Each row refers to a different sample size $n=50,100,250,500,1000$.  Each column refers to a different exposure cut-off $\pi=0.05,0.1,0.3,0.5,0.8,1$. Different types of line refer to different weight truncation (WT) strategies (solid: No WT; long-dashed: 1-99 WT; dot-dashed: 5-95 WT; dotted: 10-90 WT). The colours refer to different values of the rule-threshold $\tau$: the darker the colour, the more severe the violation (i.e., the lower $\tau$).} \label{fig:coefII:bias0}
\end{figure}
\newpage
\begin{figure}[!h]
	\includegraphics[width=0.9\textwidth]{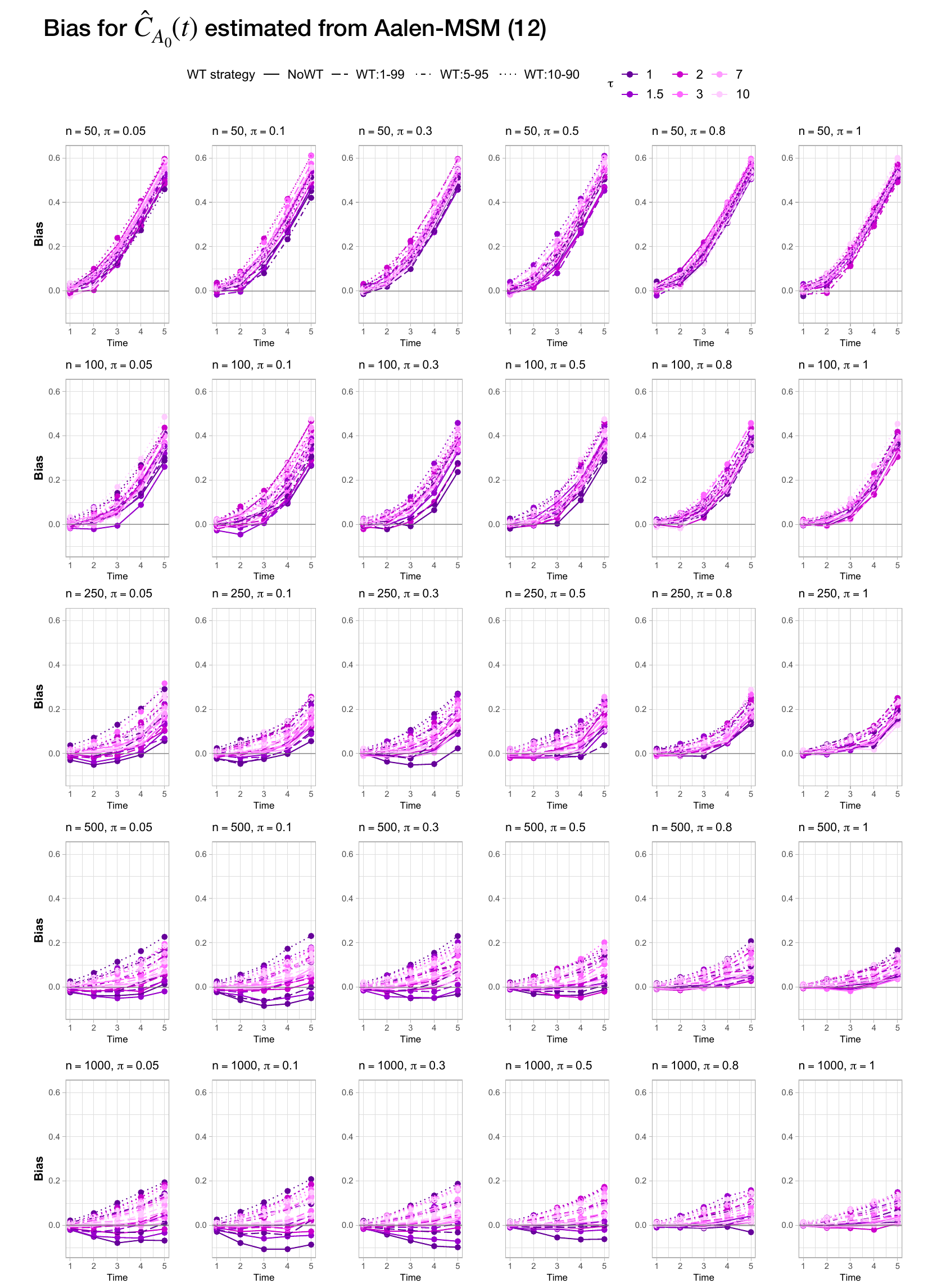}
	\caption{Bias of the estimates for the cumulative coefficient $C_{A0}(t) =\int_{0}^{t}\tilde{\alpha}_{A0}(s)ds$ at time points $t=1,\dots,5$ for the different setting of simulation study II. Each row refers to a different sample size $n=50,100,250,500,1000$.  Each column refers to a different exposure cut-off $\pi=0.05,0.1,0.3,0.5,0.8,1$. Different types of line refer to different weight truncation (WT) strategies (solid: No WT; long-dashed: 1-99 WT; dot-dashed: 5-95 WT; dotted: 10-90 WT). The colours refer to different values of the rule-threshold $\tau$: the darker the colour, the more severe the violation (i.e., the lower $\tau$).} \label{fig:coefII:biasA0}
\end{figure}
\newpage
\begin{figure}[!h]
	\includegraphics[width=0.9\textwidth]{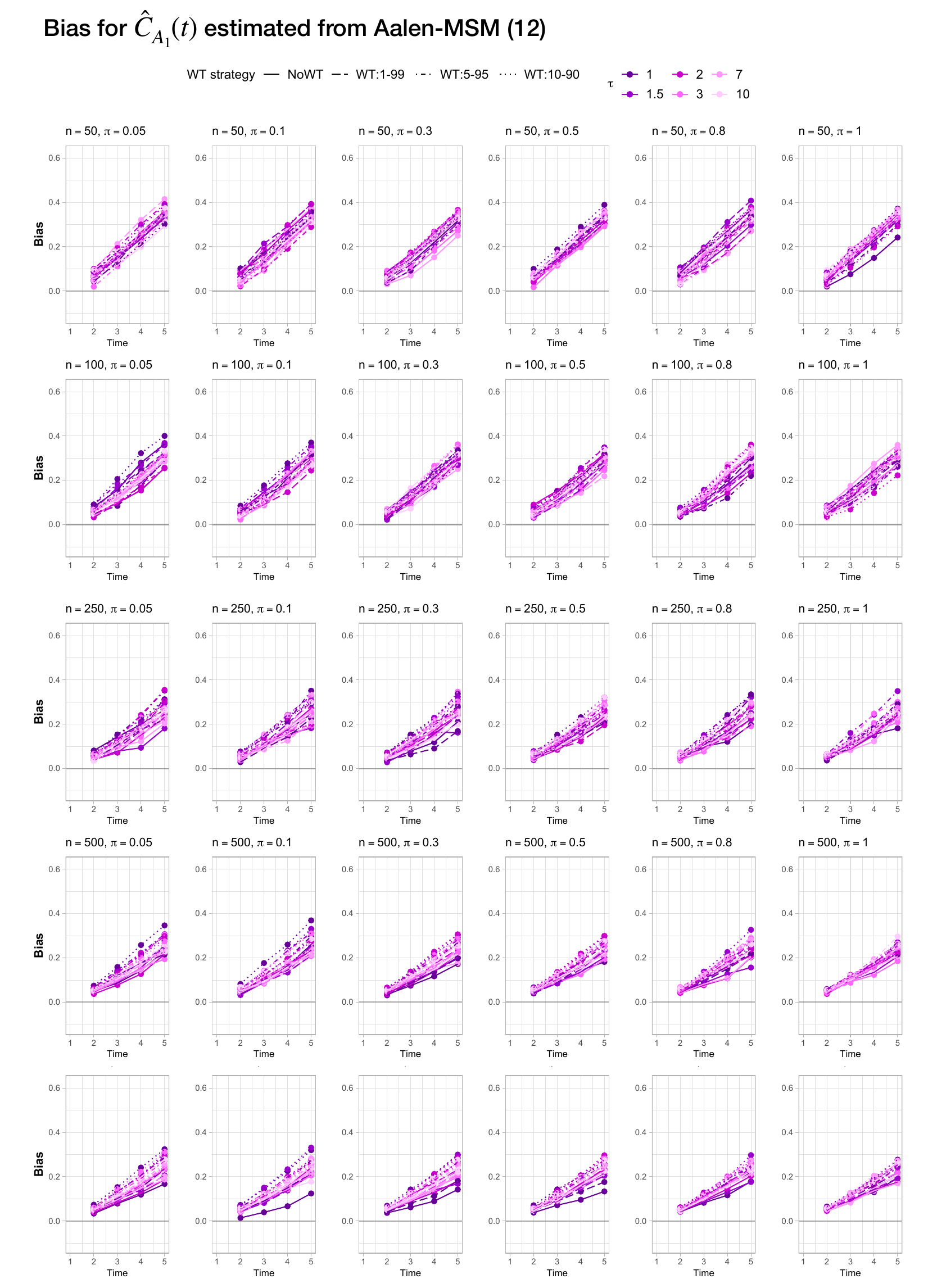}
	\caption{Bias of the estimates for the cumulative coefficient $C_{A1}(t) =\int_{1}^{t}\tilde{\alpha}_{A1}(s)ds$ at time points $t=2,3,4,5$ for the different setting of simulation study II. Each row refers to a different sample size $n=50,100,250,500,1000$.  Each column refers to a different exposure cut-off $\pi=0.05,0.1,0.3,0.5,0.8,1$. Different types of line refer to different weight truncation (WT) strategies (solid: No WT; long-dashed: 1-99 WT; dot-dashed: 5-95 WT; dotted: 10-90 WT). The colours refer to different values of the rule-threshold $\tau$: the darker the colour, the more severe the violation (i.e., the lower $\tau$).} \label{fig:coefII:biasA1}
\end{figure}
\newpage
\begin{figure}[!h]
	\includegraphics[width=0.9\textwidth]{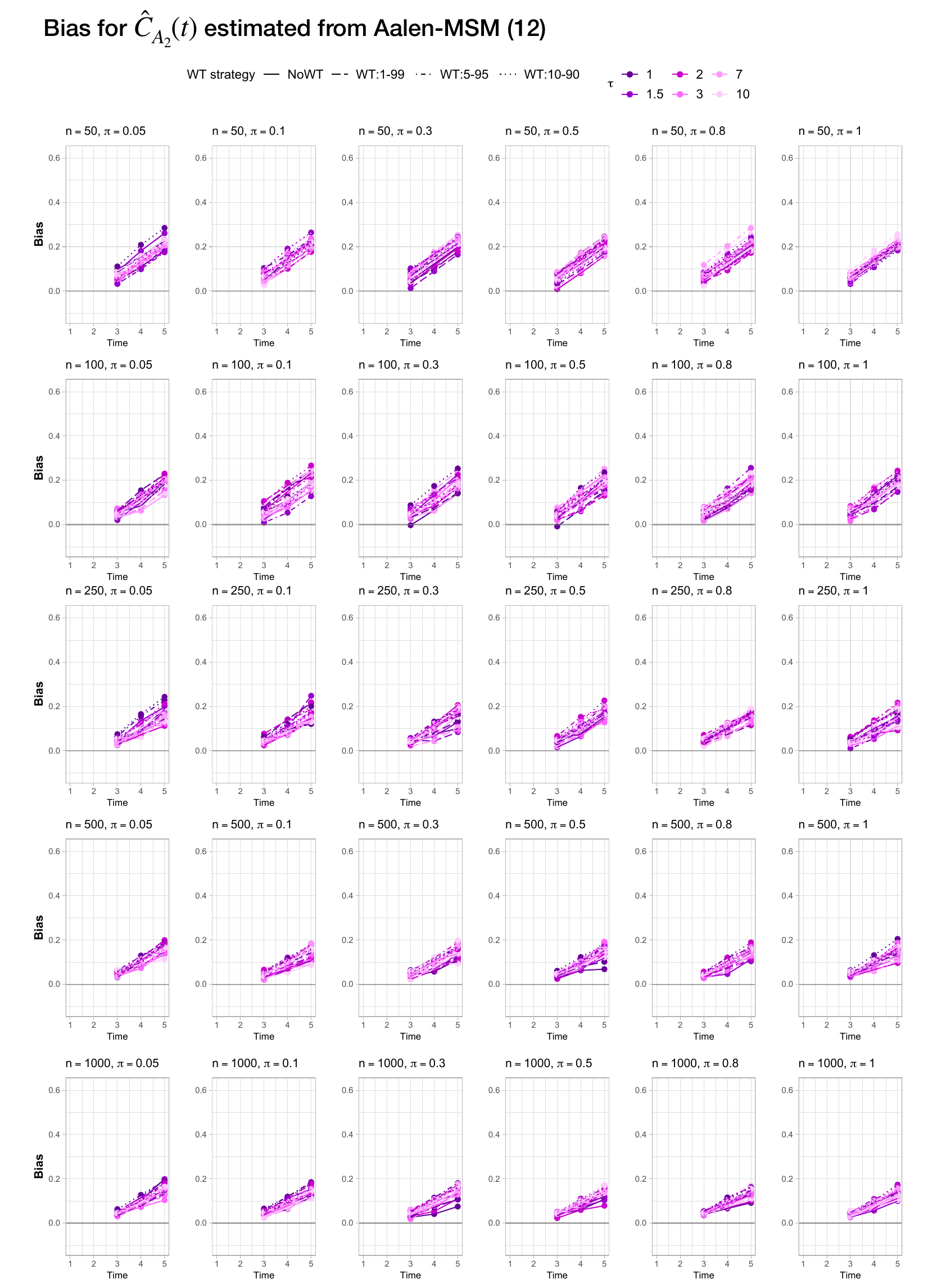}
	\caption{Bias of the estimates for the cumulative coefficient $C_{A2}(t) =\int_{2}^{t}\tilde{\alpha}_{A2}(s)ds$ at time points $t=3,4,5$ for the different setting of simulation study II. Each row refers to a different sample size $n=50,100,250,500,1000$.  Each column refers to a different exposure cut-off $\pi=0.05,0.1,0.3,0.5,0.8,1$. Different types of line refer to different weight truncation (WT) strategies (solid: No WT; long-dashed: 1-99 WT; dot-dashed: 5-95 WT; dotted: 10-90 WT). The colours refer to different values of the rule-threshold $\tau$: the darker the colour, the more severe the violation (i.e., the lower $\tau$).} \label{fig:coefII:biasA2}
\end{figure}
\newpage
\begin{figure}[!h]
	\includegraphics[width=0.9\textwidth]{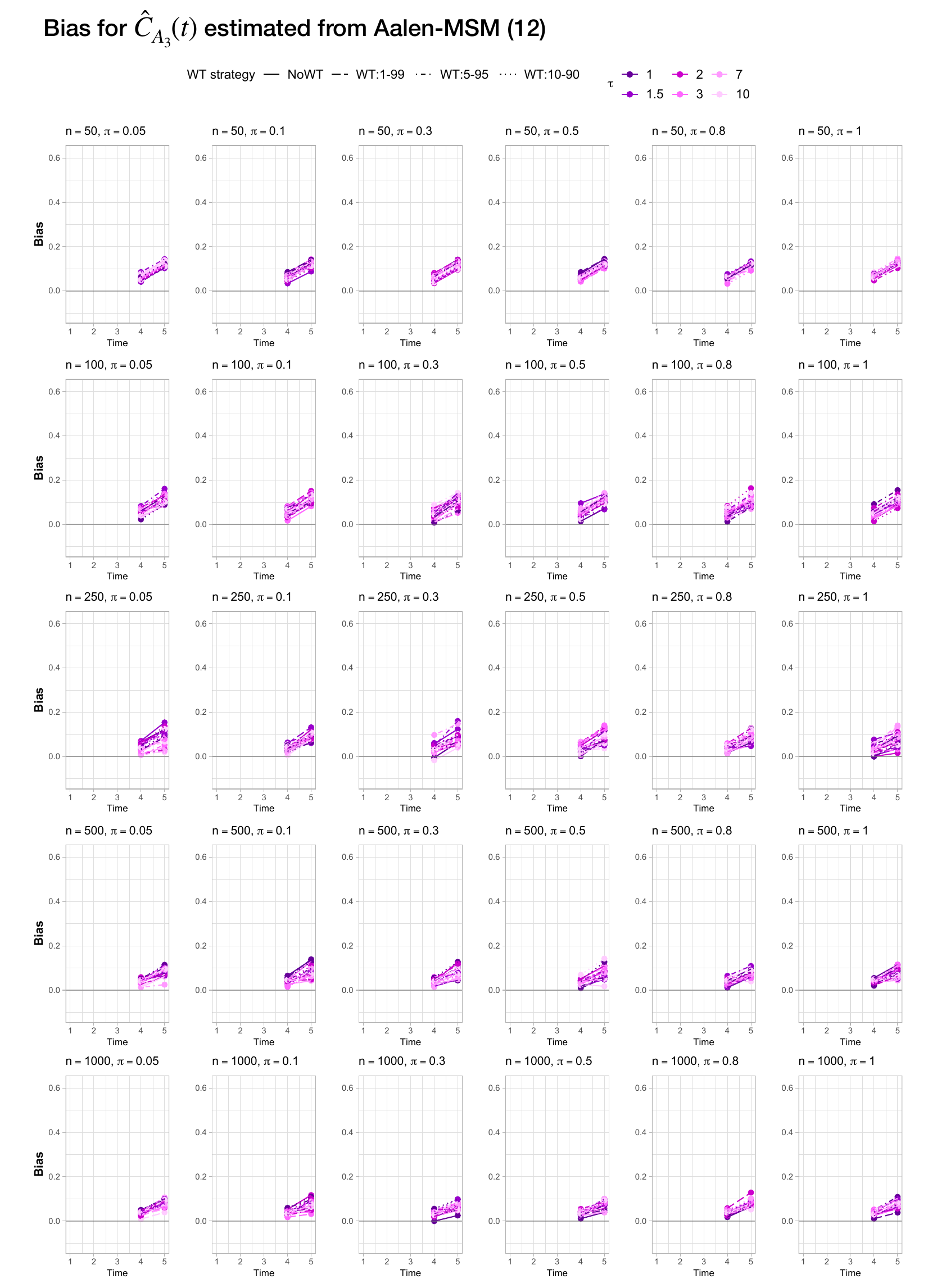}
	\caption{Bias of the estimates for the cumulative coefficient $C_{A3}(t) =\int_{3}^{t}\tilde{\alpha}_{A3}(s)ds$ at time points $t=4,5$ for the different setting of simulation study II. Each row refers to a different sample size $n=50,100,250,500,1000$.  Each column refers to a different exposure cut-off $\pi=0.05,0.1,0.3,0.5,0.8,1$. Different types of line refer to different weight truncation (WT) strategies (solid: No WT; long-dashed: 1-99 WT; dot-dashed: 5-95 WT; dotted: 10-90 WT). The colours refer to different values of the rule-threshold $\tau$: the darker the colour, the more severe the violation (i.e., the lower $\tau$).} \label{fig:coefII:biasA3}
\end{figure}
\newpage
\begin{figure}[!h]
	\includegraphics[width=0.9\textwidth]{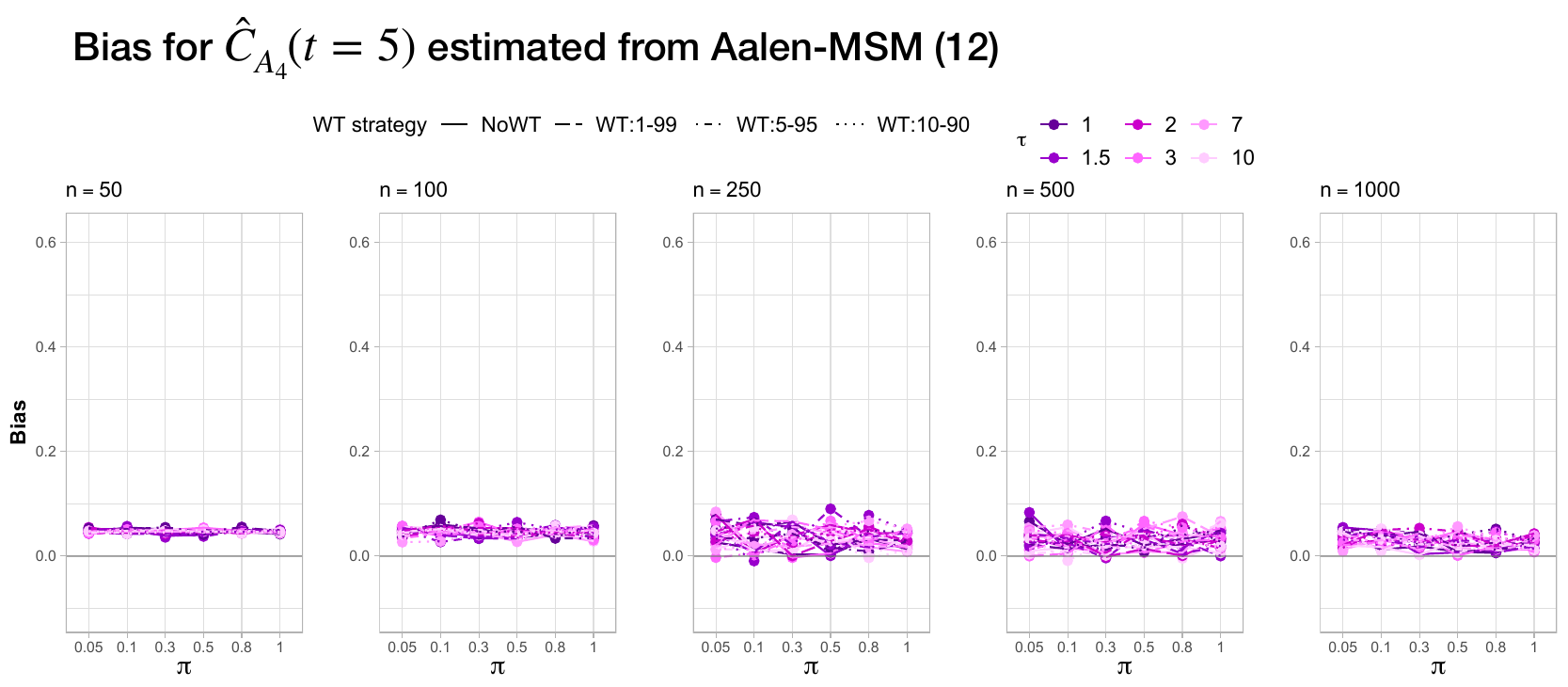}
	\caption{Bias of the estimates for the cumulative coefficient $C_{A4}(t=5) =\int_{4}^{5}\tilde{\alpha}_{A4}(s)ds$ for the different setting of simulation study II. Each column refers to a different sample size $n=50,100,250,500,1000$.  The x-axes show the compliance-threshold values  $\pi=0.05,0.1,0.3,0.5,0.8,1$. Different types of line refer to different weight truncation (WT) strategies (solid: No WT; long-dashed: 1-99 WT; dot-dashed: 5-95 WT; dotted: 10-90 WT). The colours refer to different values of the rule-threshold $\tau$: the darker the colour, the more severe the violation (i.e., the lower $\tau$).} \label{fig:coefII:biasA4}
\end{figure}

\newpage
\subsection{Empirical standard error}
\begin{figure}[!h]
	\includegraphics[width=0.9\textwidth]{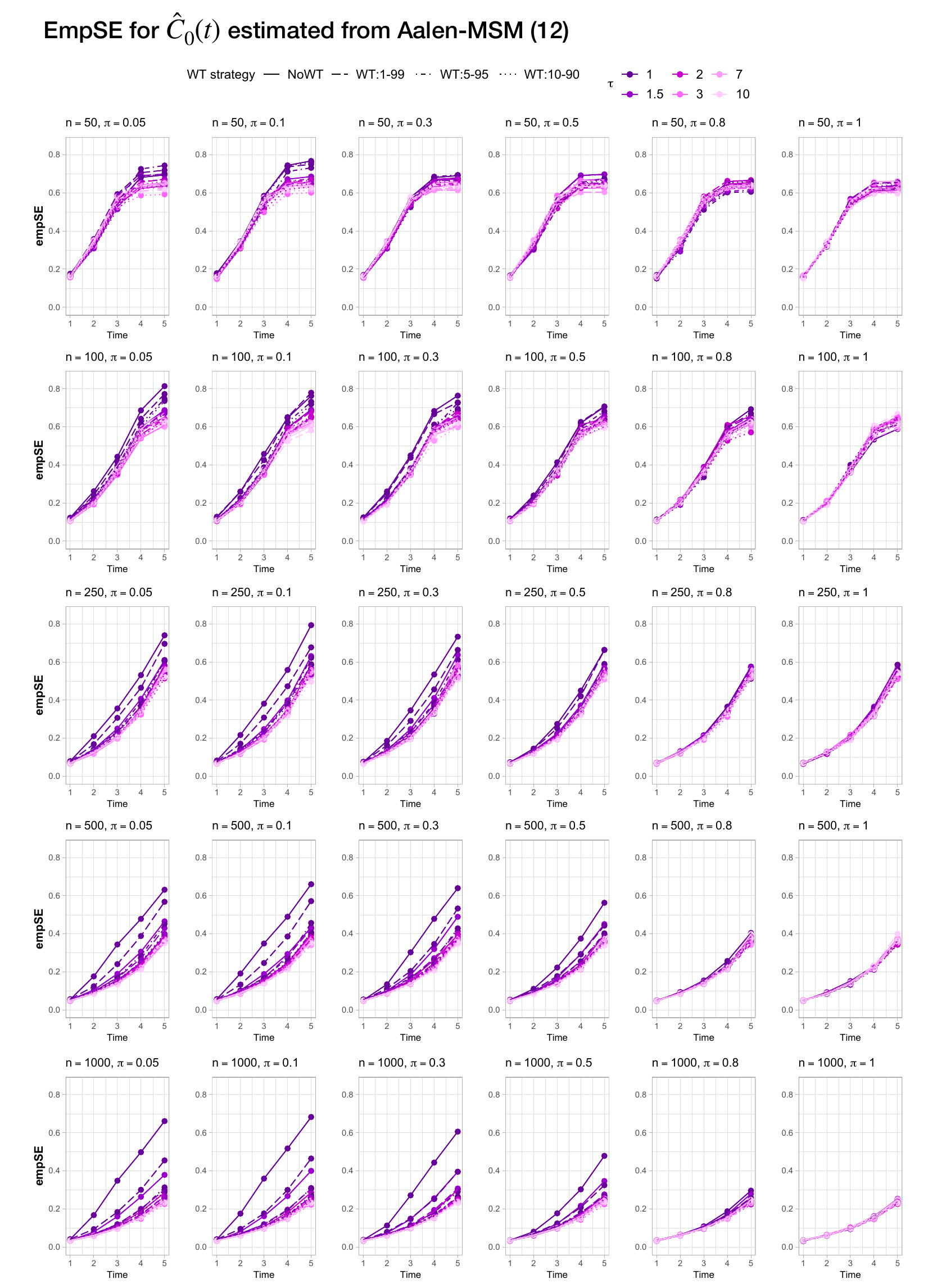}
	\caption{Empirical Standard Error (empSE) of the estimates for the cumulative coefficient $C_{0}(t) =\int_{0}^{t}\tilde{\alpha}_{0}(s)ds$ at time points $t=1,\dots,5$ for the different setting of simulation study II. Each row refers to a different sample size $n=50,100,250,500,1000$.  Each column refers to a different exposure cut-off $\pi=0.05,0.1,0.3,0.5,0.8,1$. Different types of line refer to different weight truncation (WT) strategies (solid: No WT; long-dashed: 1-99 WT; dot-dashed: 5-95 WT; dotted: 10-90 WT). The colours refer to different values of the rule-threshold $\tau$: the darker the colour, the more severe the violation (i.e., the lower $\tau$).} \label{fig:coefII:empse0}
\end{figure}
\newpage
\begin{figure}[!h]
	\includegraphics[width=0.9\textwidth]{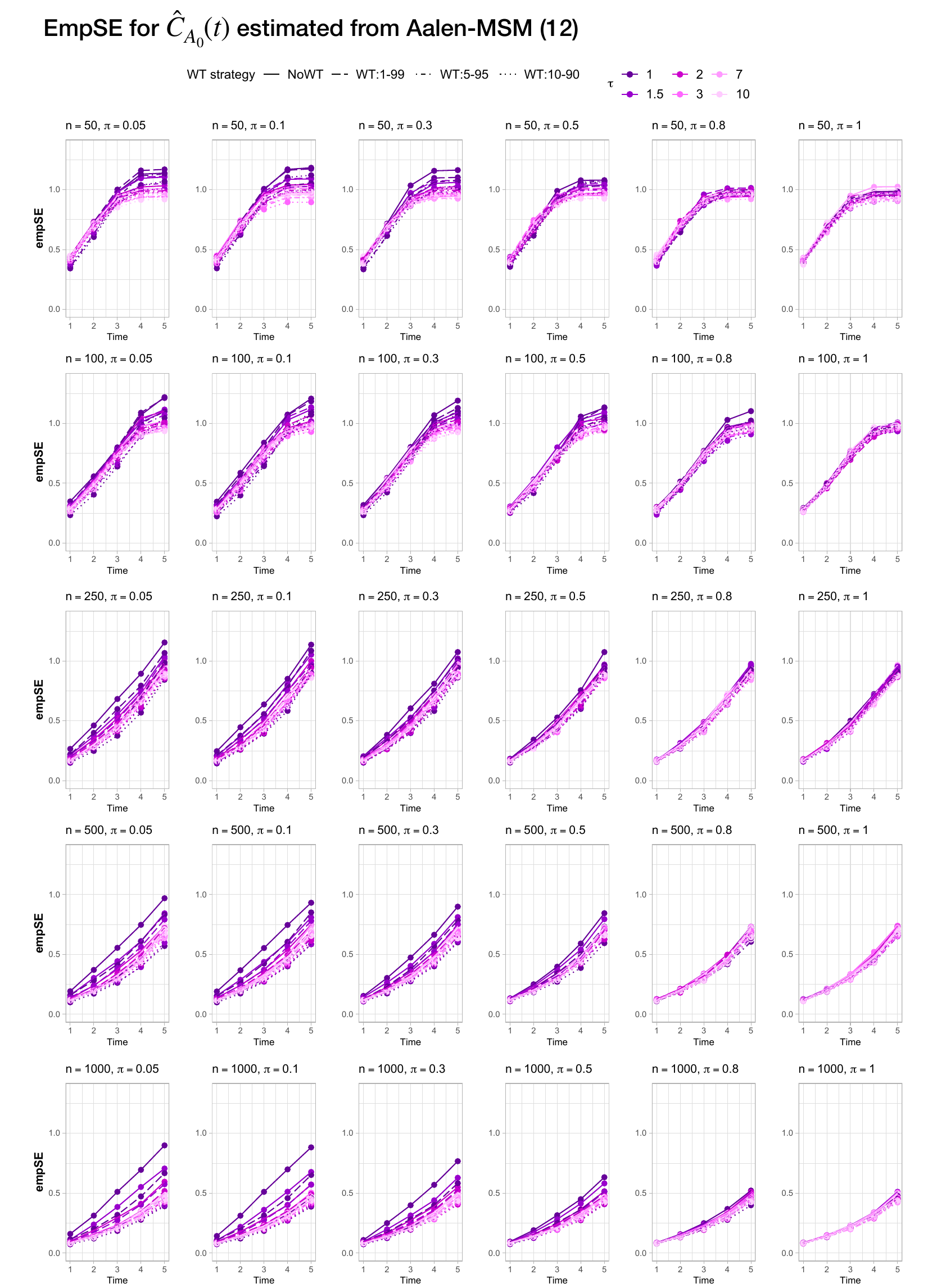}
	\caption{Empirical Standard Error (empSE) of the estimates for the cumulative coefficient $C_{A0}(t) =\int_{0}^{t}\tilde{\alpha}_{A0}(s)ds$ at time points $t=1,\dots,5$ for the different setting of simulation study II. Each row refers to a different sample size $n=50,100,250,500,1000$.  Each column refers to a different exposure cut-off $\pi=0.05,0.1,0.3,0.5,0.8,1$. Different types of line refer to different weight truncation (WT) strategies (solid: No WT; long-dashed: 1-99 WT; dot-dashed: 5-95 WT; dotted: 10-90 WT). The colours refer to different values of the rule-threshold $\tau$: the darker the colour, the more severe the violation (i.e., the lower$\tau$).} \label{fig:coefII:empseA0}
\end{figure}
\newpage
\begin{figure}[!h]
	\includegraphics[width=0.9\textwidth]{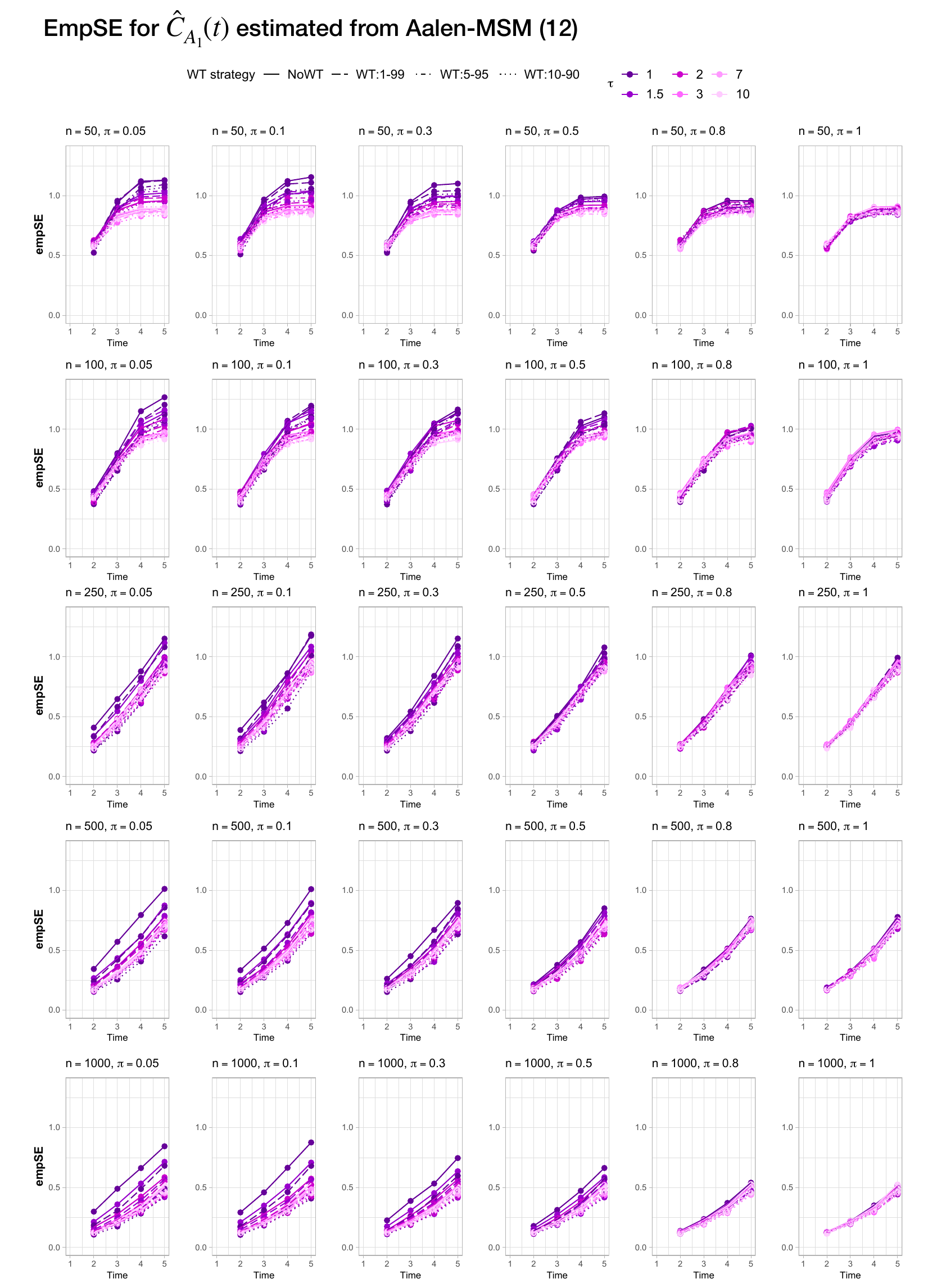}
	\caption{Empirical Standard Error (empSE)of the estimates for the cumulative coefficient $C_{A1}(t) =\int_{1}^{t}\tilde{\alpha}_{A1}(s)ds$ at time points $t=2,3,4,5$ for the different setting of simulation study II. Each row refers to a different sample size $n=50,100,250,500,1000$.  Each column refers to a different exposure cut-off $\pi=0.05,0.1,0.3,0.5,0.8,1$. Different types of line refer to different weight truncation (WT) strategies (solid: No WT; long-dashed: 1-99 WT; dot-dashed: 5-95 WT; dotted: 10-90 WT). The colours refer to different values of the rule-threshold $\tau$: the darker the colour, the more severe the violation (i.e., the lower $\tau$).} \label{fig:coefII:empseA1}
\end{figure}
\newpage
\begin{figure}[!h]
	\includegraphics[width=0.9\textwidth]{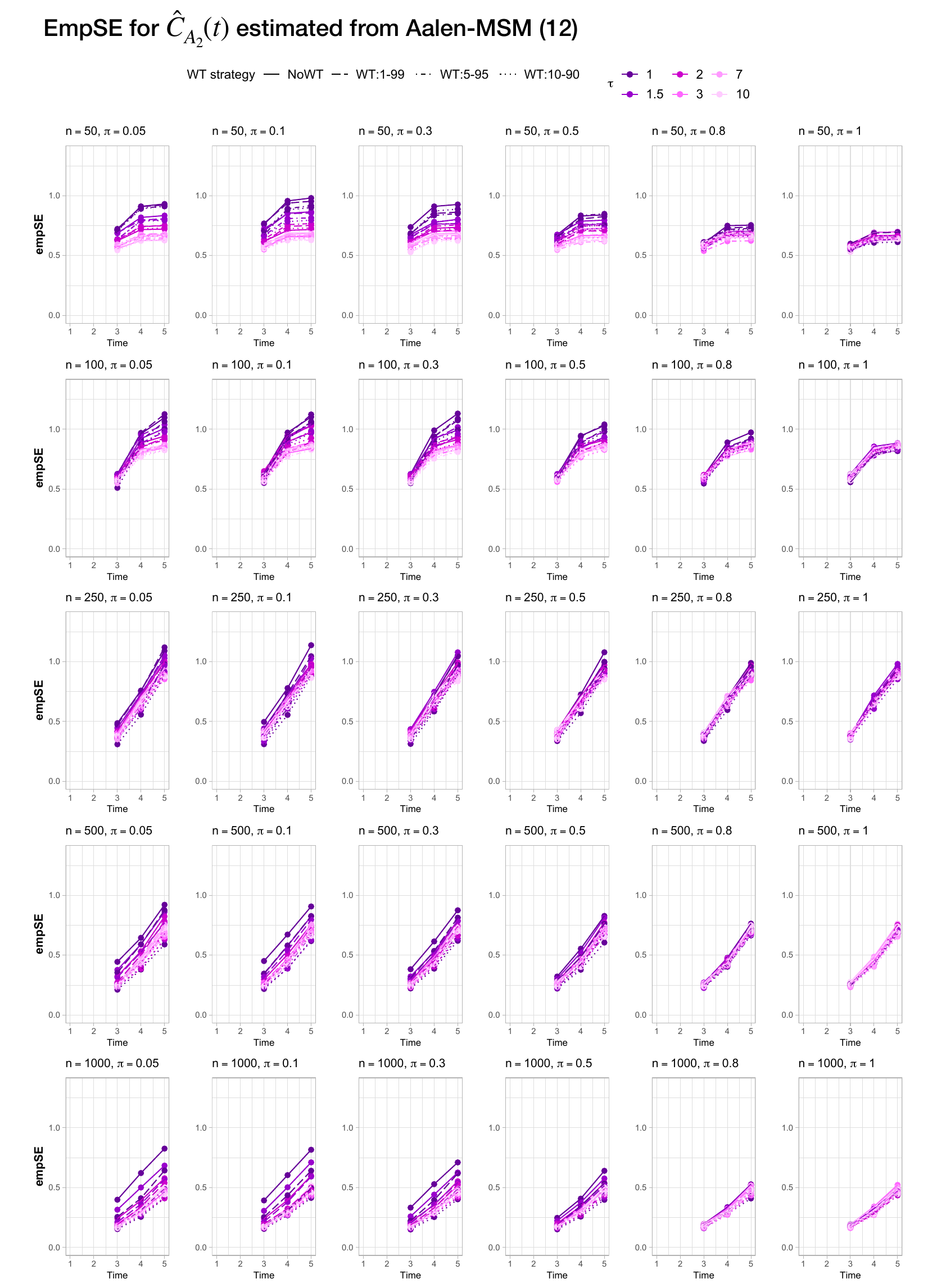}
	\caption{Empirical Standard Error (empSE)of the estimates for the cumulative coefficient $C_{A2}(t) =\int_{2}^{t}\tilde{\alpha}_{A2}(s)ds$ at time points $t=3,4,5$ for the different setting of simulation study II. Each row refers to a different sample size $n=50,100,250,500,1000$.  Each column refers to a different exposure cut-off $\pi=0.05,0.1,0.3,0.5,0.8,1$. Different types of line refer to different weight truncation (WT) strategies (solid: No WT; long-dashed: 1-99 WT; dot-dashed: 5-95 WT; dotted: 10-90 WT). The colours refer to different values of the rule-threshold $\tau$: the darker the colour, the more severe the violation (i.e., the lower $\tau$).} \label{fig:coefII:empseA2}
\end{figure}
\newpage
\begin{figure}[!h]
	\includegraphics[width=0.9\textwidth]{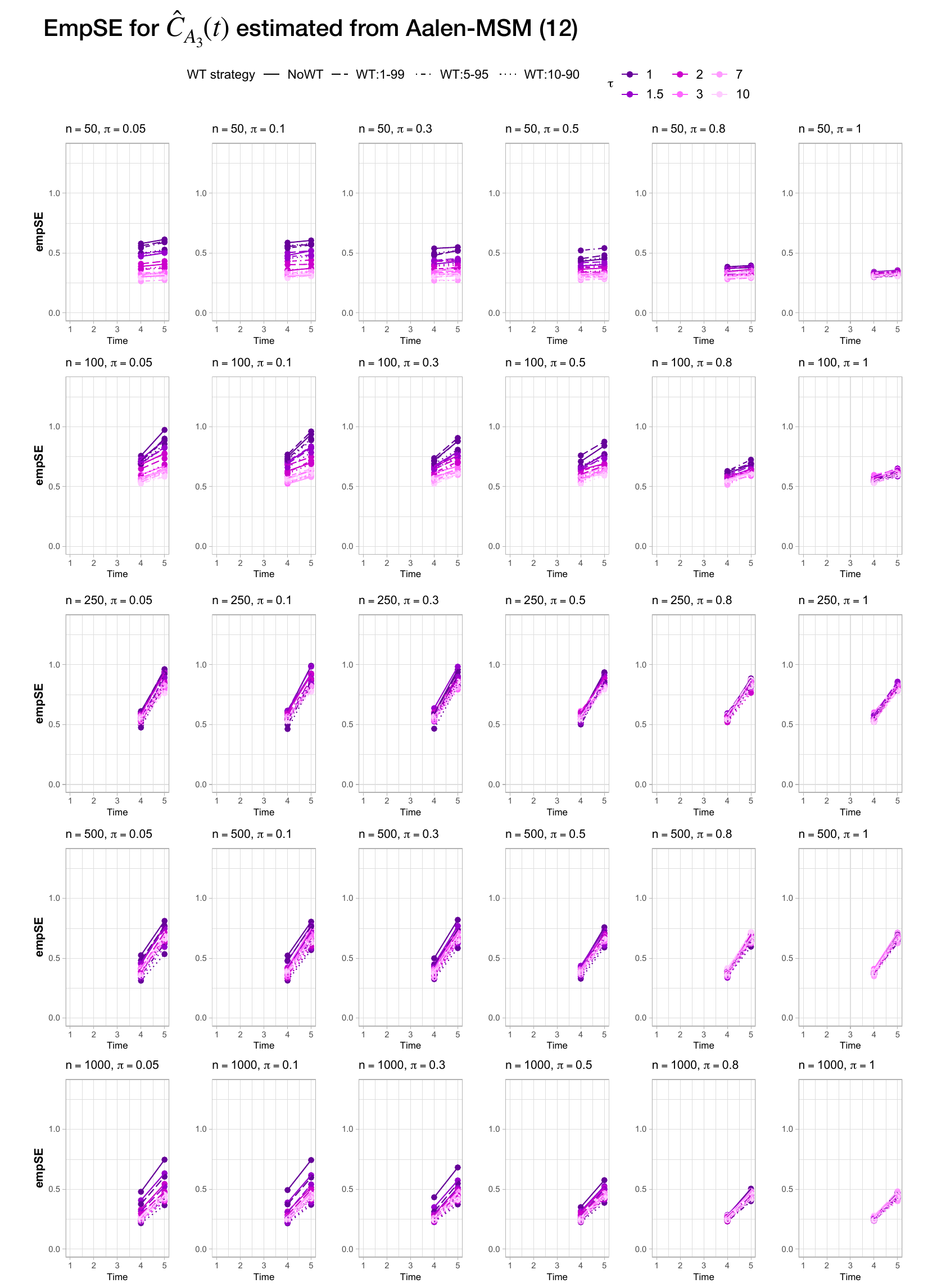}
	\caption{Empirical Standard Error (empSE)of the estimates for the cumulative coefficient $C_{A3}(t) =\int_{3}^{t}\tilde{\alpha}_{A3}(s)ds$ at time points $t=4,5$ for the different setting of simulation study II. Each row refers to a different sample size $n=50,100,250,500,1000$.  Each column refers to a different exposure cut-off $\pi=0.05,0.1,0.3,0.5,0.8,1$. Different types of line refer to different weight truncation (WT) strategies (solid: No WT; long-dashed: 1-99 WT; dot-dashed: 5-95 WT; dotted: 10-90 WT). The colours refer to different values of the rule-threshold $\tau$: the darker the colour, the more severe the violation (i.e., the lower $\tau$).} \label{fig:coefII:empseA3}
\end{figure}
\newpage
\begin{figure}[!h]
	\includegraphics[width=0.9\textwidth]{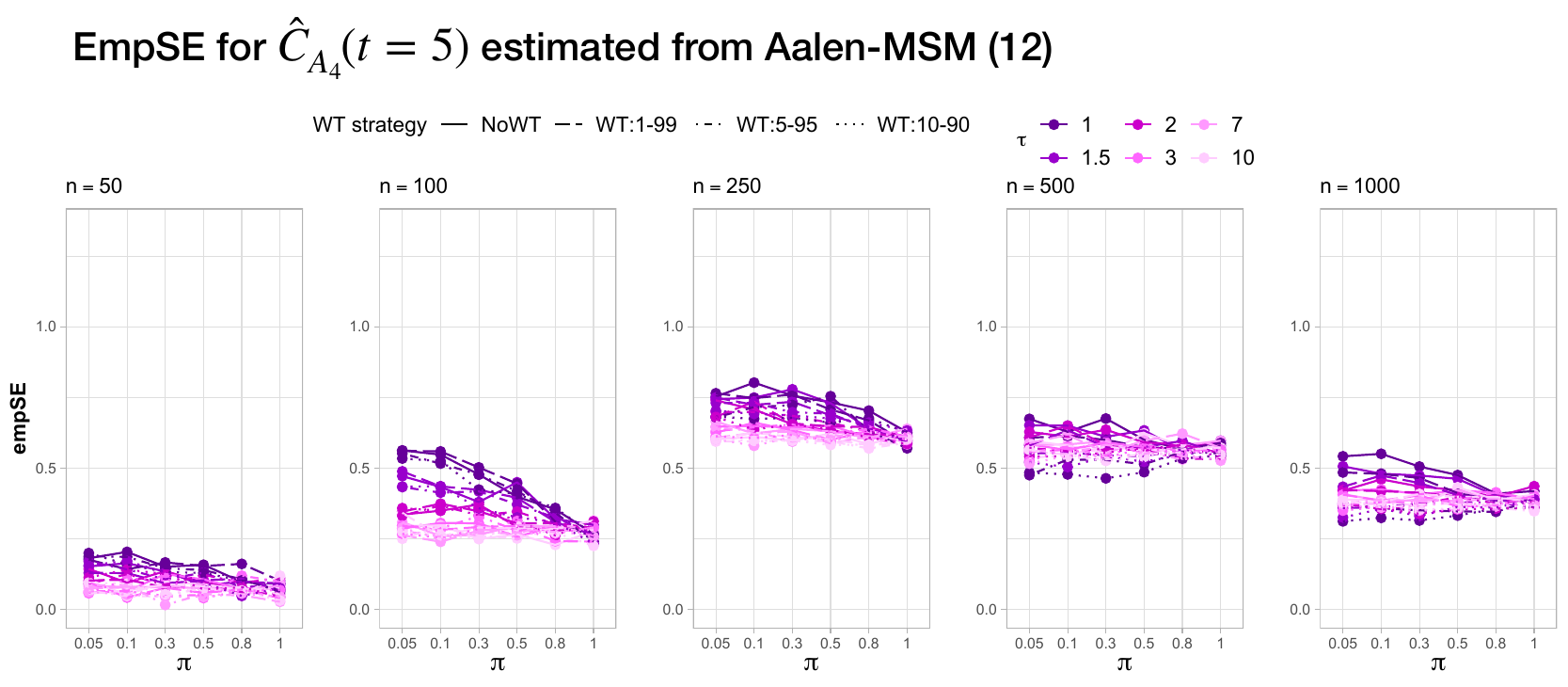}
	\caption{Empirical Standard Error (empSE) of the estimates for the cumulative coefficient $C_{A4}(t=5) =\int_{4}^{5}\tilde{\alpha}_{A4}(s)ds$ for the different setting of simulation study II. Each column refers to a different sample size $n=50,100,250,500,1000$.  The x-axes show the compliance-threshold values  $\pi=0.05,0.1,0.3,0.5,0.8,1$. Different types of line refer to different weight truncation (WT) strategies (solid: No WT; long-dashed: 1-99 WT; dot-dashed: 5-95 WT; dotted: 10-90 WT). The colours refer to different values of the rule-threshold $\tau$: the darker the colour, the more severe the violation (i.e., the lower $\tau$).} \label{fig:coefII:empseA4}
\end{figure}

\newpage
\subsection{Root mean squared error }
\begin{figure}[!h]
	\includegraphics[width=0.9\textwidth]{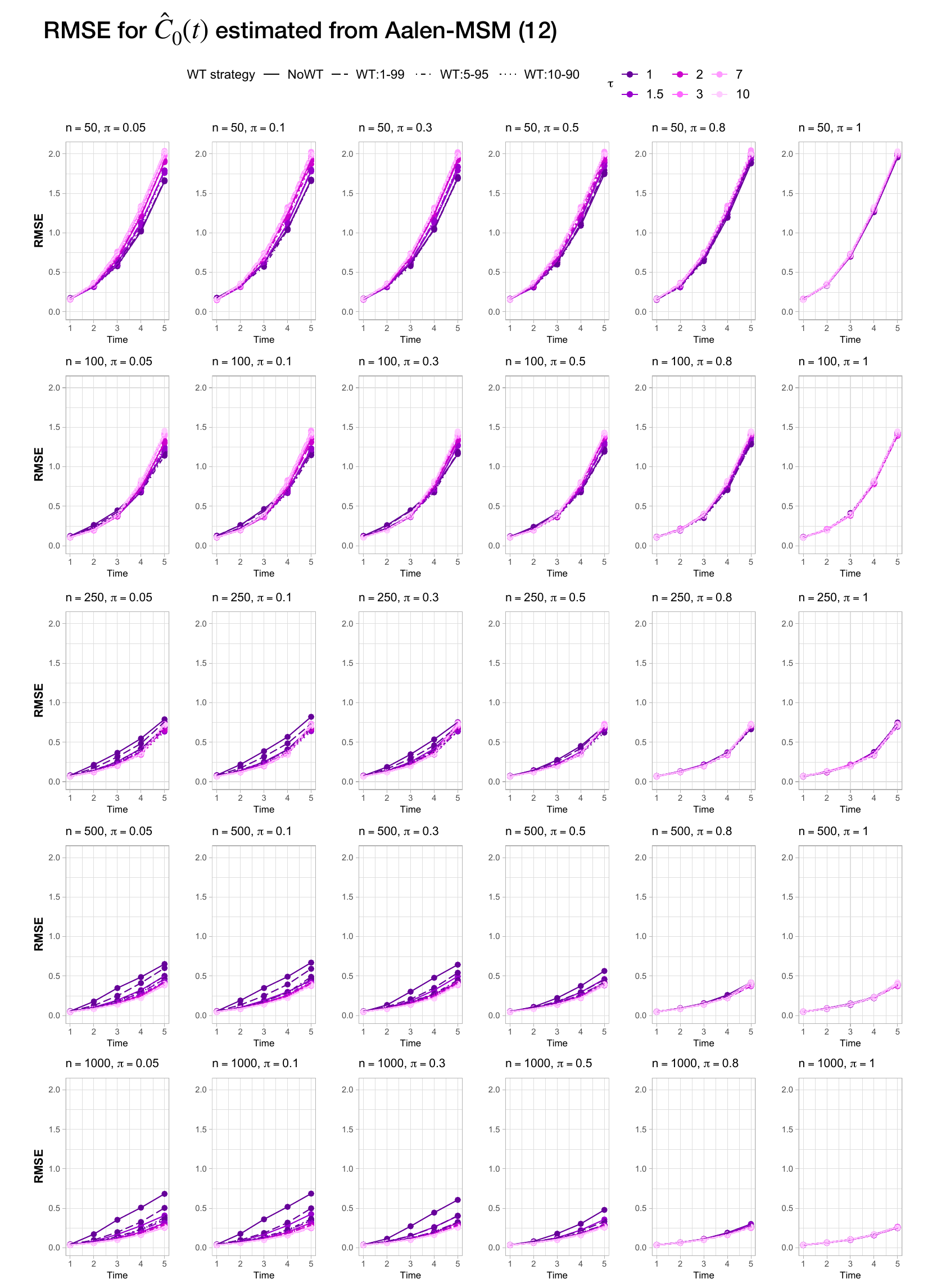}
	\caption{Root Mean Squared Error (RMSE) of the estimates for the cumulative coefficient $C_{0}(t) =\int_{0}^{t}\tilde{\alpha}_{0}(s)ds$ at time points $t=1,\dots,5$ for the different setting of simulation study II. Each row refers to a different sample size $n=50,100,250,500,1000$.  Each column refers to a different exposure cut-off $\pi=0.05,0.1,0.3,0.5,0.8,1$. Different types of line refer to different weight truncation (WT) strategies (solid: No WT; long-dashed: 1-99 WT; dot-dashed: 5-95 WT; dotted: 10-90 WT). The colours refer to different values of the rule-threshold $\tau$: the darker the colour, the more severe the violation (i.e., the lower $\tau$).} \label{fig:coefII:rmse0}
\end{figure}
\newpage
\begin{figure}[!h]
	\includegraphics[width=0.9\textwidth]{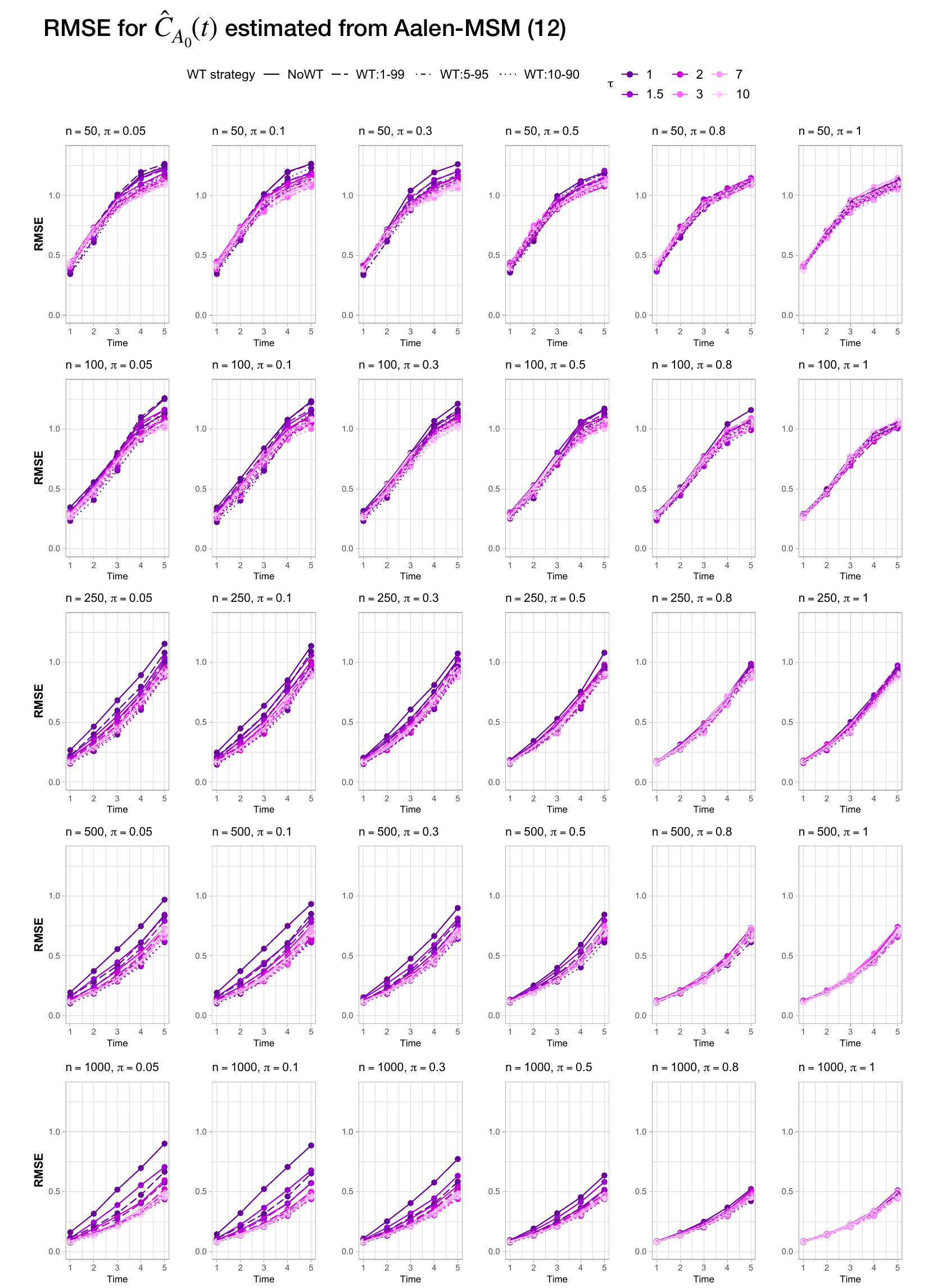}
	\caption{Root Mean Squared Error (RMSE) of the estimates for the cumulative coefficient $C_{A0}(t) =\int_{0}^{t}\tilde{\alpha}_{A0}(s)ds$ at time points $t=1,\dots,5$ for the different setting of simulation study II. Each row refers to a different sample size $n=50,100,250,500,1000$.  Each column refers to a different exposure cut-off $\pi=0.05,0.1,0.3,0.5,0.8,1$. Different types of line refer to different weight truncation (WT) strategies (solid: No WT; long-dashed: 1-99 WT; dot-dashed: 5-95 WT; dotted: 10-90 WT). The colours refer to different values of the rule-threshold $\tau$: the darker the colour, the more severe the violation (i.e., the lower $\tau$).} \label{fig:coefII:rmseA0}
\end{figure}
\newpage
\begin{figure}[!h]
	\includegraphics[width=0.9\textwidth]{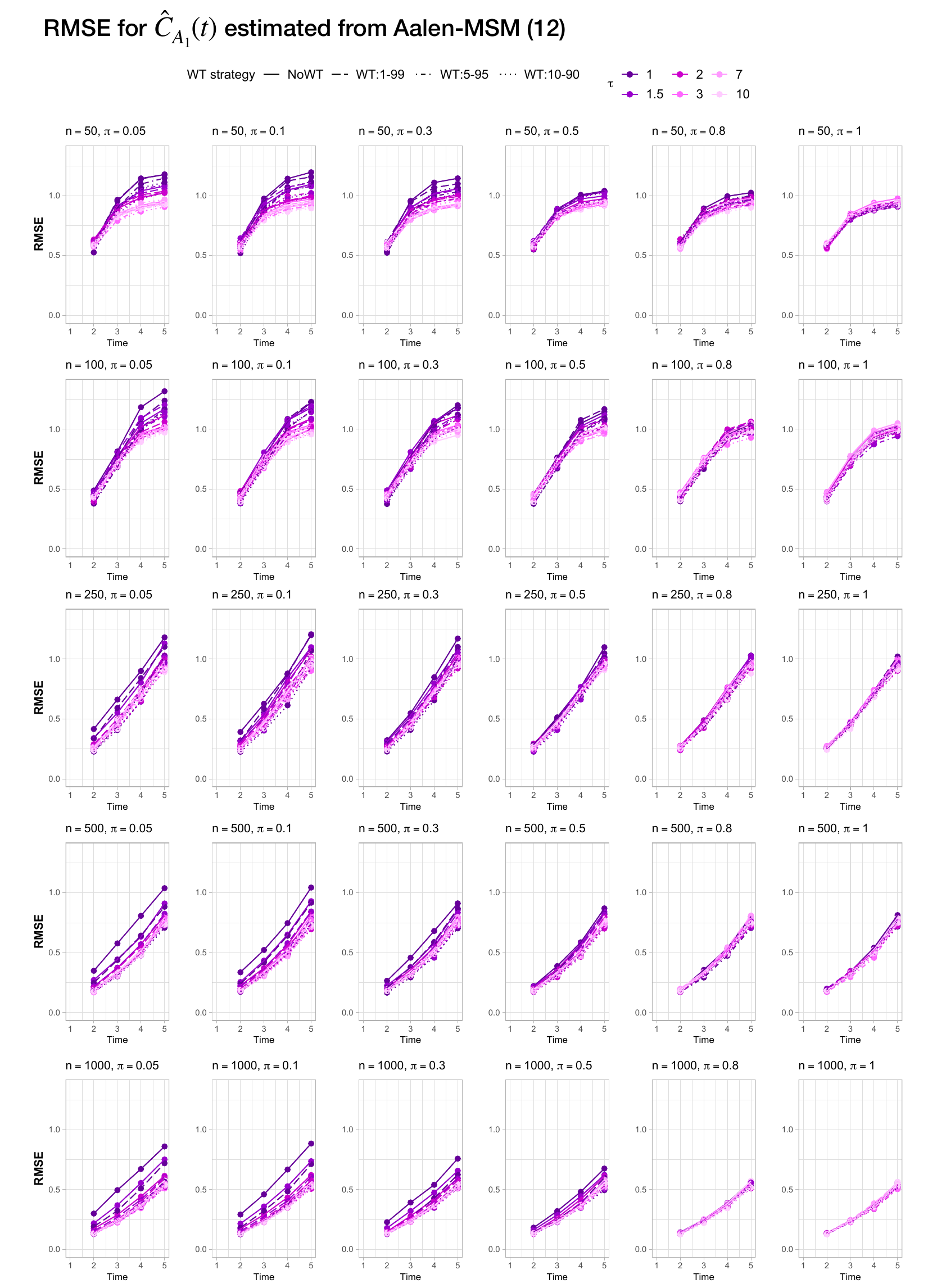}
	\caption{Root Mean Squared Error (RMSE) of the estimates for the cumulative coefficient $C_{A1}(t) =\int_{1}^{t}\tilde{\alpha}_{A1}(s)ds$ at time points $t=2,3,4,5$ for the different setting of simulation study II. Each row refers to a different sample size $n=50,100,250,500,1000$.  Each column refers to a different exposure cut-off $\pi=0.05,0.1,0.3,0.5,0.8,1$. Different types of line refer to different weight truncation (WT) strategies (solid: No WT; long-dashed: 1-99 WT; dot-dashed: 5-95 WT; dotted: 10-90 WT). The colours refer to different values of the rule-threshold $\tau$: the darker the colour, the more severe the violation (i.e., the lower $\tau$).} \label{fig:coefII:rmseA1}
\end{figure}
\newpage
\begin{figure}[!h]
	\includegraphics[width=0.9\textwidth]{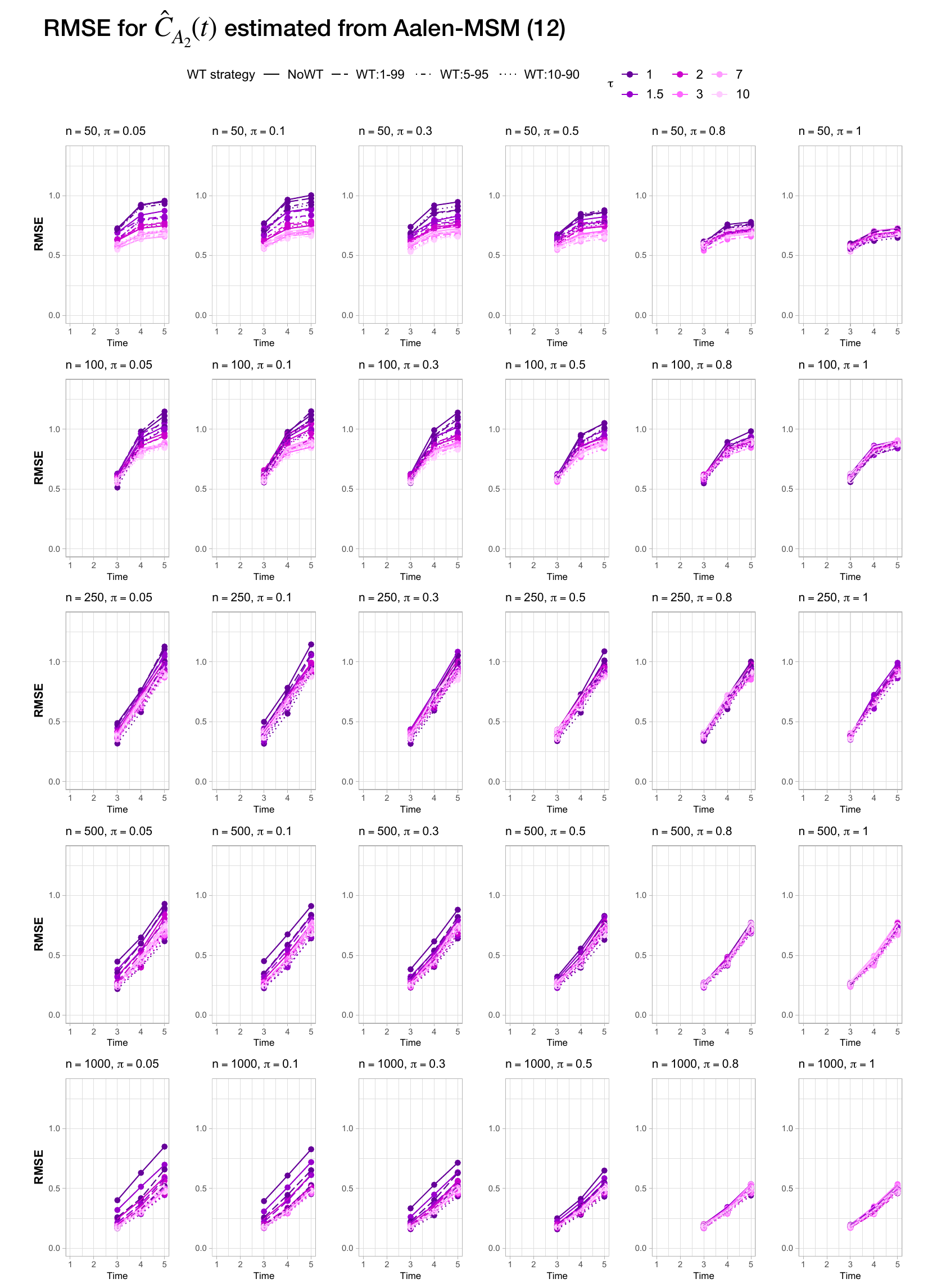}
	\caption{Root Mean Squared Error (RMSE)of the estimates for the cumulative coefficient $C_{A2}(t) =\int_{2}^{t}\tilde{\alpha}_{A2}(s)ds$ at time points $t=3,4,5$ for the different setting of simulation study II. Each row refers to a different sample size $n=50,100,250,500,1000$.  Each column refers to a different exposure cut-off $\pi=0.05,0.1,0.3,0.5,0.8,1$. Different types of line refer to different weight truncation (WT) strategies (solid: No WT; long-dashed: 1-99 WT; dot-dashed: 5-95 WT; dotted: 10-90 WT). The colours refer to different values of the rule-threshold $\tau$: the darker the colour, the more severe the violation (i.e., the lower $\tau$).} \label{fig:coefII:rmseA2}
\end{figure}
\newpage
\begin{figure}[!h]
	\includegraphics[width=0.9\textwidth]{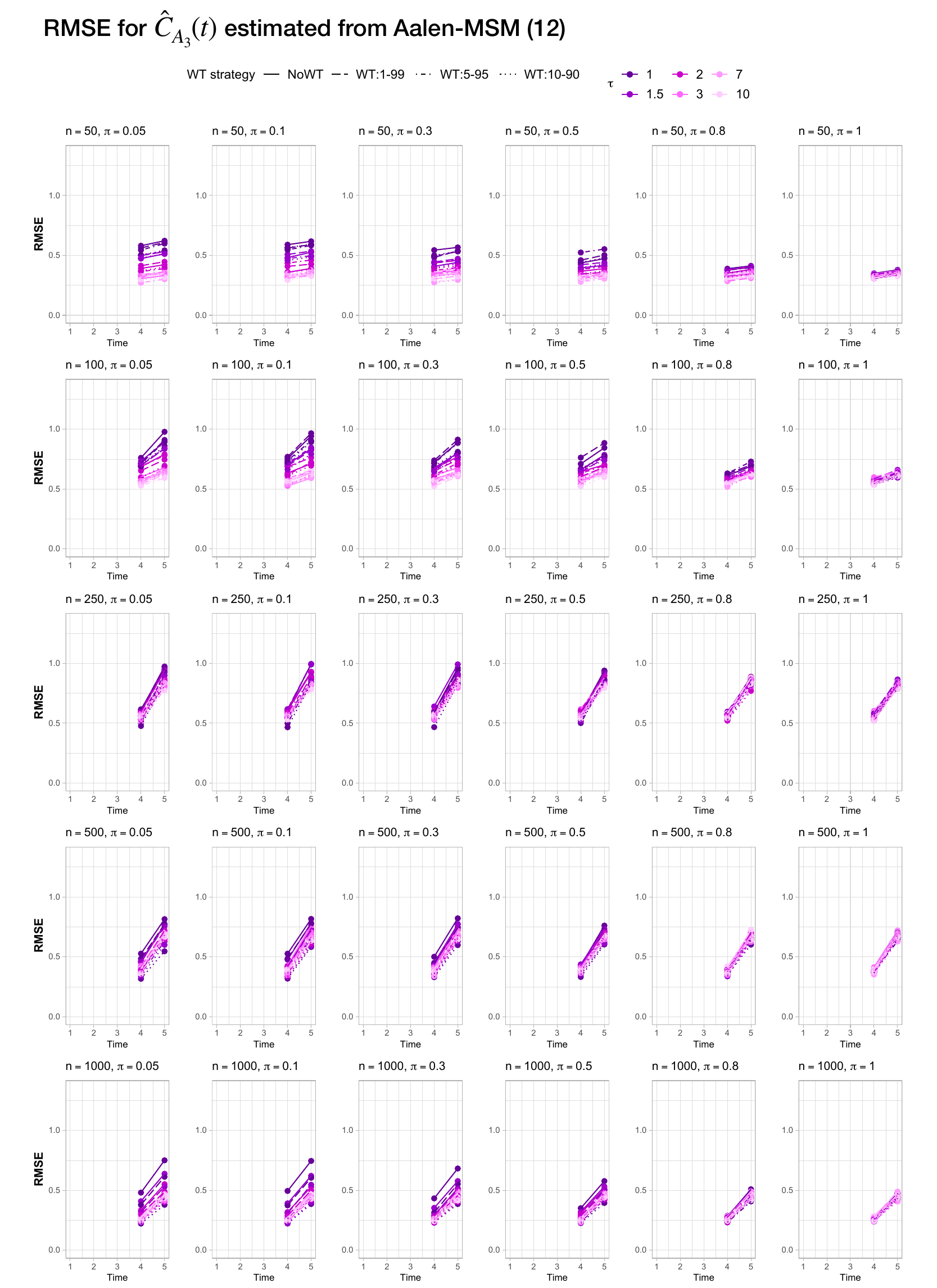}
	\caption{Root Mean Squared Error (RMSE)of the estimates for the cumulative coefficient $C_{A3}(t) =\int_{3}^{t}\tilde{\alpha}_{A3}(s)ds$ at time points $t=4,5$ for the different setting of simulation study II. Each row refers to a different sample size $n=50,100,250,500,1000$.  Each column refers to a different exposure cut-off $\pi=0.05,0.1,0.3,0.5,0.8,1$. Different types of line refer to different weight truncation (WT) strategies (solid: No WT; long-dashed: 1-99 WT; dot-dashed: 5-95 WT; dotted: 10-90 WT). The colours refer to different values of the rule-threshold $\tau$: the darker the colour, the more severe the violation (i.e., the lower $\tau$).} \label{fig:coefII:rmseA3}
\end{figure}
\newpage
\begin{figure}[!h]
	\includegraphics[width=0.9\textwidth]{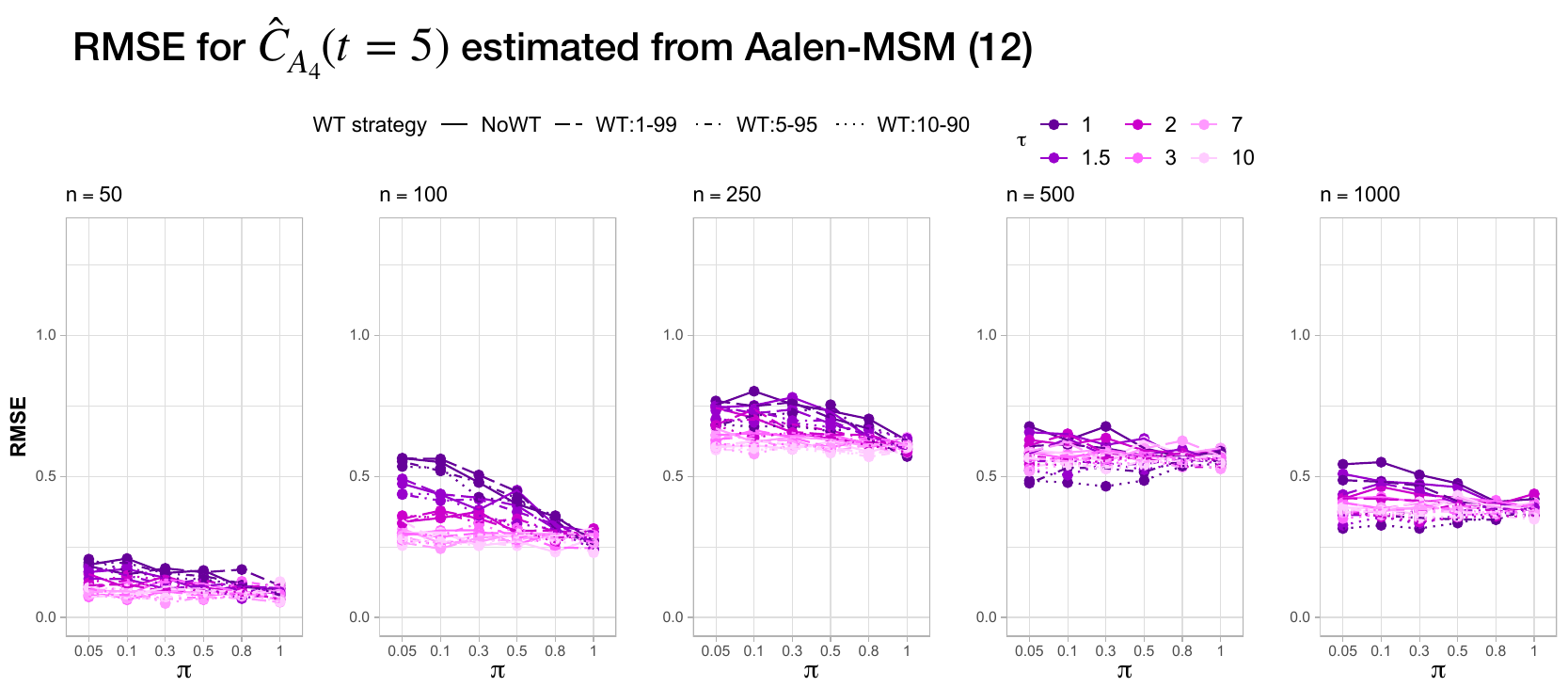}
	\caption{Root Mean Squared Error (RMSE) of the estimates for the cumulative coefficient $C_{A4}(t=5) =\int_{4}^{5}\tilde{\alpha}_{A4}(s)ds$ for the different setting of simulation study II. Each column refers to a different sample size $n=50,100,250,500,1000$.  The x-axes show the compliance-threshold values  $\pi=0.05,0.1,0.3,0.5,0.8,1$. Different types of line refer to different weight truncation (WT) strategies (solid: No WT; long-dashed: 1-99 WT; dot-dashed: 5-95 WT; dotted: 10-90 WT). The colours refer to different values of the rule-threshold $\tau$: the darker the colour, the more severe the violation (i.e., the lower $\tau$).} \label{fig:coefII:rmseA4}
\end{figure}

\end{document}